\documentclass[12pt]{article}

\usepackage{amsmath,amssymb,amsfonts,amsthm,hyperref}
\usepackage{graphicx,multirow,caption,bbm,latexsym,epsfig}
\usepackage{cite}
\usepackage{physics}
\usepackage{feynmp-auto}
\usepackage{cancel}

\newcommand{\bs}{\begin{subequations}}
\newcommand{\es}{\end{subequations}}
\newcommand{\be}{\begin{equation}}
\newcommand{\ee}{\end{equation}}
\newcommand{\ba}{\begin{eqnarray}}
\newcommand{\ea}{\end{eqnarray}}
\newcommand{\no}{\nonumber \\}

\newcommand{\AAA}{\mathcal{A}}
\newcommand{\SSS}{\mathcal{S}}
\newcommand{\RRR}{\mathcal{R}}

\newcommand{\marrow}[6]{
    \fmfcmd{style_def marrow#1
    expr p = drawarrow subpath #6 of p shifted 10 #2 withpen pencircle scaled 0.5; label.#3(btex #4 etex, point 0.5 of p shifted 12 #2);
    enddef;}
    \fmf{marrow#1,tension=0}{#5}}
    
\newcommand{\marrowii}[6]{
\fmfcmd{style_def marrow#1
expr p = drawarrow subpath #6 of p shifted 28 #2 withpen pencircle scaled 0.5; label.#3(btex #4 etex, point 0.5 of p shifted 30 #2);
enddef;}
\fmf{marrow#1,tension=0}{#5}}

\allowdisplaybreaks

\begin{document}

\begin{fmffile}{diagram}

\fmfcmd{
style_def arrow_right expr p =
  shrink(1);
    cfill(harrow (reverse p, .62));
  endshrink
enddef;
style_def wiggly_arrow expr p =
cdraw (wiggly p);
shrink (1);
cfill (arrow p);
endshrink;
enddef;} 

\title{
  \normalsize \hfill CFTP/21-013 \\[3mm]
  \LARGE
  Prescription for finite oblique parameters $S$ and $U$ \\
  in extensions of the SM with $m_W \neq m_Z \cos{\theta_W}$}

\author{Francisco~Albergaria\thanks{E-mail:
    \tt \href{mailto:francisco.albergaria@tecnico.ulisboa.pt}{francisco.albergaria@tecnico.ulisboa.pt}}
  \ and Lu\'\i s~Lavoura\thanks{E-mail:
    \tt \href{mailto:balio@cftp.tecnico.ulisboa.pt}{balio@cftp.tecnico.ulisboa.pt}}
  \\*[3mm]
  \small Universidade de Lisboa, Instituto Superior T\'ecnico, CFTP, \\
  \small Av.~Rovisco~Pais~1, 1049-001~Lisboa, Portugal}

\maketitle

\begin{abstract}
  We consider extensions of the Standard Model
  with neutral scalars in multiplets of $SU(2)$ larger than doublets.
  When those scalars acquire vacuum expectation values,
  the resulting masses of the gauge bosons $W^\pm$ and $Z^0$
  are not related by $m_W = m_Z \cos{\theta_W}$.
  In those extensions of the Standard Model
  the oblique parameters $S$ and $U$,
  when computed at the one-loop level,
  turn out to be either gauge-dependent or divergent.
  We show that one may eliminate this problem
  by modifying the Feynman rules of the Standard Model
  for some vertices containing the Higgs boson;
  the modifying factors are equal to $1$
  in the limit $m_W = m_Z \cos{\theta_W}$.
  We give the result for $S$ in a model with arbitrary numbers
  of scalar $SU(2)$ triplets with weak hypercharges either $0$ or $1$.
\end{abstract}

\newpage

\section{Introduction}
\label{introduction}

The present knowledge of particle physics is encapsulated
in the Standard Model (SM)~\cite{Glashow1961,Weinberg1967,Salam1968}.
This is an $SU(3) \times SU(2) \times U(1)$ gauge theory
that describes all the fundamental particles observed until now
and the way they interact with each other.
The scalar sector of the SM contains just one doublet of $SU(2)$;
it is responsible for giving masses $m_W$ and $m_Z$
to the gauge bosons $W^\pm$ and $Z^0$,
respectively,
as well as to the fermions.

Despite being one of the most accurate theories in Science,
there are phenomena that the SM cannot explain.
Therefore particle physicists try to either extend or complete it.
One of the ways to extend the SM is
by enlarging its scalar sector~\cite{Ivanov2017}.
The most studied extension of that sector
is the addition of another $SU(2)$ doublet,
obtaining a two-Higgs-doublet model (2HDM)~\cite{Branco2012}.
Extensions of the SM with scalar $SU(2)$ singlets
are also frequent in the literature
(see for instance Refs.~\cite{Profumo2007,Ahriche2014,Costa2017}).
There are also some extensions of the SM
with scalar $SU(2)$ triplets in the literature
(see for instance Refs.~\cite{Arhrib2011,Arhrib2012,Kanemura20121,Kanemura20122,Kanemura2013,Xu2016,Azevedo2021,Chiang2021,Ait-Ouazghour2021}).

A feature of the SM is custodial symmetry.
This is a symmetry of the scalar potential of the SM
that is broken both by the Yukawa interactions
and by the gauge coupling of the $U(1)$ group.
Custodial symmetry leads at the tree level to the relation
\be
\label{relation}
m_W = m_Z c_W,
\ee
where $c_W$ is the cosine of the Weinberg angle.
The relation~\eqref{relation} is in good agreement with observation:
experimental results give~\cite{Zyla2020}
\be
\rho \equiv \frac{m_W^2}{m_Z^2 c_W^2} = 1.00038 \pm 0.00020.
\ee
Models in which only scalar $SU(2)$ singlets and doublets
have vacuum expectation values (VEVs)
preserve the relation~\eqref{relation} at the tree level.
That relation is broken at the loop level
because the custodial symmetry is not exact,
as noted above.

Peskin and Takeuchi have identified three
so-called `oblique parameters' that they named $S$,
$T$,
and $U$~\cite{Peskin1990,Peskin1992}.\footnote{At about the same time,
Altarelli and Barbieri
  defined equivalent parameters $\epsilon_1 = \alpha T$,
  $\epsilon_2 = - \alpha U \left/ \left( 4 s_W^2 \right) \right.$,
  and $\epsilon_3 = \alpha S \left/ \left( 4 s_W^2 \right) \right.$,
  where $\alpha$ is the fine-structure constant and $s_W$ is the sine
  of the Weinberg angle~\cite{Altarelli1991,Altarelli1992}.}
These three observables parameterize some effects of New Physics (NP),
\textit{i.e.}\ physics of extensions of the SM.
At the one-loop level,
vacuum polarization produces the tensors $\Pi_{V V^\prime}^{\mu \nu} (q)$,
where $V$ and $V^\prime$ are gauge bosons that may be either
$A$ and $A$ ($A$ stands for the photon),
$A$ and $Z$,
$Z$ and $Z$,
or $W$ and $W$,
and $q$ is the four-momentum of the gauge bosons.
We write those tensors as
\be
\Pi_{V V^\prime}^{\mu \nu} (q)
= g^{\mu \nu} A_{V V^\prime} (q^2) + q^\mu q^\nu B_{V V^\prime} (q^2).
\ee
We then define
$\delta A_{V V^\prime} (q^2) \equiv \left. A_{V V^\prime}(q^2) \right|_\mathrm{NP}
- \left. A_{V V^\prime}(q^2) \right|_\mathrm{SM}$,
where $\left. A_{V V^\prime}(q^2) \right|_\mathrm{NP}$ is $A_{V V^\prime}(q^2)$
computed in a given extension of the SM
and $\left. A_{V V^\prime}(q^2) \right|_\mathrm{SM}$
is $A_{V V^\prime}(q^2)$ computed in the SM.
The oblique parameters are defined as
\bs
\ba
T &=& \frac{1}{\alpha m_Z^2}
\left[ \frac{\delta A_{WW}(0)}{c_W^2} - \delta A_{ZZ}(0) \right],
\label{eq:T}\\
S  &=& \frac{4 s_W^2 c_W^2}{\alpha}
\left[
  \left. \frac{\partial \, \delta A_{Z Z}(q^2)}{\partial q^2} \right|_{q^2=0}
  - \left. \frac{\partial \, \delta A_{A A}(q^2)}{\partial q^2} \right|_{q^2=0}
  + \frac{c_W^2 - s_W^2}{c_W s_W}
  \left. \frac{\partial \, \delta A_{A Z}(q^2)}{\partial q^2} \right|_{q^2=0}
  \right],
\label{eq:S} \\
U  &=& \frac{4 s_W^2}{\alpha}
\left[ \left. \frac{\partial \, \delta A_{W W}(q^2)}{\partial q^2} \right|_{q^2=0}
  - c_W^2\,
  \left. \frac{\partial \, \delta A_{Z Z}(q^2)}{\partial q^2} \right|_{q^2=0}
  \right. \no & & \left.
  - s_W^2\,
  \left.
  \frac{\partial \, \delta A_{A A}(q^2)}{\partial q^2} \right|_{q^2=0}
  + 2 c_W s_W\,
  \left. \frac{\partial \, \delta A_{A Z} (q^2)}{\partial q^2} \right|_{q^2=0}
  \right].
\label{eq:U}
\ea
\es

The oblique parameters have already been computed
for several extensions of the SM,
in particular for models with arbitrary numbers of scalar doublets
and singlets~\cite{Grimus2008,Grimus20082},
wherein they are finite.
In models where
larger scalar multiplets are added to the SM and have VEVs,
$m_W$ is in general different from $m_Z c_W$.
As we will show in this paper,
this gives rise to complications
when computing the oblique parameters.\footnote{More generally,
models with $m_W \neq m_Z c_W$
lead to various complications in their renormalization,
as emphasized in Ref.~\cite{glusza}.
This is because in those models $\rho$ is an additional parameter
that needs to be independently renormalized.
See Refs.~\cite{jegerlehner,lynn,blank} too.}
First of all,
the oblique parameter $T$ is divergent at one-loop level
for models that do not have custodial symmetry~\cite{Chanowitz1985,Gunion1991}.
Furthermore,
as noted above,
when computing the oblique parameters one needs to subtract
the functions $A_{V V^\prime} (q^2)$ computed in the SM
from the same functions computed in the extension of the SM.
This subtraction is non-trivial,
since in the SM the masses of the gauge bosons obey equation~\eqref{relation}
while in the extension they do not obey it.
This leads to \emph{divergent} parameters $S$ and $U$.
We have found out that,
in order to obtain \emph{finite}
(and gauge-independent)
results for $S$ and $U$ at the one-loop level in models with $m_W \neq m_Z c_W$,
one needs to multiply the usual SM Feynman rules\footnote{The Feyman rules
  for the SM may be found,
  for instance,
  in Refs.~\cite{Branco1999} and~\cite{Romao2012}.}
for some vertices containing the SM Higgs boson
by factors that become equal to $1$ when $m_W = m_Z c_W$
but are different from $1$ for $m_W \neq m_Z c_W$.

This paper is organized as follows.
In section~\ref{sec:general} we consider an extension of the SM
with arbitrary numbers of arbitrarily large
$SU(2) \times U(1)$ scalar multiplets,
subject only to the constraint of the preservation of electric charge.
In section~\ref{sec:prescription}
we illustrate the problem that arises
in the computation of both $S$ and $U$ in that extension of the SM,
and display our proposed solution to that problem.
In section~\ref{sec:triplets} we restrict the model of section~\ref{sec:general}
to the case where only scalar $SU(2)$ triplets and singlets
with $U(1)$ charges 1 or 0 are added to the SM;
we give the explicit expression for $S$ in that model.
Section~\ref{sec:conclusions} summarizes our findings.
Several appendices contain technical material
that may be avoided by a hurried reader.

\section{General extension of the SM}
\label{sec:general}

We consider an $SU(2) \times U(1)$ electroweak model
where the scalar sector consists of an arbitrary number of multiplets $M_{JY}$
labeled by their weak isospin $J$
and weak hypercharge $Y$.
We restrict ourselves to models where all the scalars
have integer electric charges\footnote{In our normalization of $Y$,
the electric charge $Q$ is given by $Q = T_3 + Y$,
where $T_3$ is the third component of weak isospin.};
therefore,
if $J$ is a (half-)integer then $Y$ is a (half-)integer too.
Furthermore,
we consider only complex multiplets;
if some multiplets are real,
then our conclusions are still valid
but there are some modifications in the intermediate steps.
Finally,
for the sake of simplicity we assume that there is only one $M_{JY}$
for each value of the pair $\left( J,\, Y \right)$;
our conclusions are still valid otherwise,
but the notation would become clumsier.

Each multiplet $M_{JY}$ has a component $M_{JY}^Q$ with electric charge $Q$
if and only if $J - \left| Q - Y \right| \in \mathbb{N}_0$.
Some multiplets $M_{JY}$ may have,
for some values of $Q \neq 0$,
both a component $M_{JY}^Q$ with electric charge $Q$
and a component $M_{JY}^{-Q}$ with electric charge $-Q$;
in our notation,
$M_{JY}^Q \neq \left( M_{JY}^{-Q} \right)^\ast$.

The covariant derivative for the $SU(2) \times U(1)$ electroweak model is,
as usual,\footnote{We use the sign conventions of Ref.~\cite{Branco1999}.
They correspond to setting $\eta = - 1$ and $\eta_Z = \eta_e = 1$
in Ref.~\cite{Romao2012}.}
\be
\label{covderiv}
D_\mu = \partial_\mu + i e Q A_\mu
- i\, \frac{g}{c_W} \left( T_3 - Q s_W^2 \right) Z_\mu
- i g \left( W_\mu^+ T_+ + W_\mu^- T_- \right),
\ee
where $e = g s_W$ is the unit of electric charge,
the $T_i$
($i = 1, 2, 3$)
are the three components of weak isospin,
and the $T_\pm$ are given by
$T_{\pm} = \left( T_1 \pm i T_2 \right) \left/ \sqrt{2} \right.$.
We use the $SU(2)$ representation for weak isospin $J$:
\bs
\allowdisplaybreaks
\label{su2}
\ba
\left( T_3 \right)_{rc} &=& \delta_{rc} \left( J + 1 - r \right),
\label{bvjfigtof}
\\
\left( T_+ \right)_{rc} &=& \delta_{r+1,c}\,
\sqrt{\frac{r\, \left( 2 J + 1 - r \right)}{2}},
\\
\left( T_- \right)_{rc} &=& \delta_{r-1,c}\,
\sqrt{\frac{\left( r - 1 \right) \left( 2 J + 2 - r \right)}{2}},
\ea
\es
where $r$ stands for the row of the matrix
and $c$ stands for the column of the matrix,
with $1 \le r, c \le 2 J + 1$.
We write $M_{JY}$ as a column matrix with $2 J + 1$ rows
upon which the matrices of Eqs.~\eqref{su2} act.
Clearly,
from Eq.~\eqref{bvjfigtof},
$M_{JY}^Q$ is in the row $J + 1 - T_3 = J + Y + 1 - Q$ of that column matrix.
Then,
using Eqs.~\eqref{covderiv} and~\eqref{su2},
\bs
\allowdisplaybreaks
\label{viw00e}
\ba
D_\mu M_{JY}^Q &=& \partial_\mu M_{JY}^Q + i e Q A_\mu M_{JY}^Q
+ i\, \frac{g}{c_w}\, Z_\mu M_{JY}^Q \left( Y - Q c_W^2 \right)
\label{u1} \\ & &
- i g\, W_\mu^+ M_{JY}^{Q-1}\,
\sqrt{\frac{\left( J + Y + 1 - Q \right) \left( J - Y + Q \right)}{2}}
\label{u2} \\ & &
- i g\, W_\mu^- M_{JY}^{Q+1}\,
\sqrt{\frac{\left( J + Y - Q \right) \left( J - Y + Q +1 \right)}{2}}.
\label{u3}
\ea
\es

We define $\AAA$ to be the set of multiplets
that have a component with zero electric charge.
A multiplet $M_{JY}$ belongs to $\AAA$
if and only if $J - \left| Y \right| \in \mathbb{N}_0$.

For every $Q \in \mathbb{N}$,
we define two sets:
\begin{itemize}
\item $\RRR_Q$ is the set of multiplets
  that have a component with electric charge $Q$;
  a multiplet $M_{JY}$ belongs to $\RRR_Q$
  if and only if $J - \left| Q - Y \right| \in \mathbb{N}_0$.
\item $\SSS_Q$ is the set of multiplets
  that have a component with electric charge $-Q$;
  a multiplet $M_{JY}$ belongs to $\SSS_Q$
  if and only if $J - \left| Q + Y \right| \in \mathbb{N}_0$.
\end{itemize}
We emphasize that both $\RRR_Q$ and $\SSS_Q$ are defined only for $Q > 0$.
For those $Q$,
let $n_Q$ denote the total number of charge-$Q$ scalars and let $S_a^Q$
($a=1, \dots, n_Q$)
denote the physical
(mass eigenstate)
charge-$Q$ scalars.
For definiteness,
we fix $S_1^1 \equiv G^+$ to be the charged Goldstone boson.
We define $S_a^{-Q} \equiv \left( S_a^Q \right)^\ast$.

For every $Q \in \mathbb{N}$ and $M_{JY} \in \RRR_Q$,
we write
\be
\label{8594543}
M_{JY}^Q = R_{JY}^Q
\left( \begin{array}{c} S_1^Q \\*[0.5mm] S_2^Q \\*[0.5mm] \vdots \\*[0.5mm]
  S_{n_Q}^Q \end{array} \right),
\ee
where $R_{JY}^Q$ is a $1 \times n_Q$ row matrix.
For every $Q \in \mathbb{N}$ and $M_{JY} \in \SSS_Q$,
we write
\be
\label{85945432}
\left( M_{JY}^{-Q} \right)^\ast = S_{JY}^Q
\left( \begin{array}{c} S_1^Q \\*[0.5mm] S_2^Q \\*[0.5mm] \vdots \\*[0.5mm]
  S_{n_Q}^Q \end{array} \right),
\ee
where $S_{JY}^Q$ is a $1 \times n_Q$ row matrix.
We form the $n_Q \times n_Q$ unitary (mixing) matrix $U^Q$
by piling up all the row matrices $R_{JY}^Q$ and $S_{JY}^Q$
(for fixed $Q \in \mathbb{N}$)
on top of each other.
Since $U^Q$ is unitary,
\bs
\ba
\sum_{M_{JY} \in \RRR_Q}
\left( R^Q_{JY} \right)_{1a}  \left( R^Q_{JY} \right)^\ast_{1a^\prime}
+ \sum_{M_{JY} \in \SSS_Q}
\left( S^Q_{JY} \right)_{1a}  \left( S^Q_{JY} \right)^\ast_{1a^\prime}
&=& \delta_{a a^\prime},
\label{unitarity1} \\
\sum_{a=1}^{n_Q} \left( R^Q_{JY} \right)_{1a}
\left( R^Q_{J^\prime Y^\prime} \right)_{1a}^\ast
&=& \delta_{J J^\prime} \delta_{Y Y^\prime},
\label{unitarity2} \\
\sum_{a=1}^{n_Q} \left( S^Q_{JY} \right)_{1a}
\left( S^Q_{J^\prime Y^\prime} \right)_{1a}^\ast
&=& \delta_{J J^\prime} \delta_{Y Y^\prime},
\label{unitarity3} \\
\sum_{a=1}^{n_Q} \left( R^Q_{JY} \right)_{1a}
\left( S^Q_{J^\prime Y^\prime} \right)_{1a}^\ast &=& 0.
\label{unitarity4}
\ea
\es

Each multiplet $M_{JY}$ that belongs to $\AAA$ has VEV $v_{JY}$
(which may in some cases be zero)
in its component $M_{JY}^0$ with electric charge zero.
We write
\be
\label{ure954}
M_{JY}^0 = v_{JY} + \frac{A_{JY} + i B_{JY}}{\sqrt{2}}
\left( \begin{array}{c} S_1^0 \\*[0.5mm] S_2^0 \\*[0.5mm] \vdots \\*[0.5mm]
  S_{n_0}^0 \end{array} \right),
\ee
where $A_{JY}$ and $B_{JY}$ are \emph{real} $1 \times n_0$ row matrices
and the \emph{real} fields $S_b^0$ ($b = 1, \ldots, n_0$)
are the physical neutral scalars.
Without lack of generality,
we fix $S_1^0 \equiv G^0$ to be the neutral Goldstone boson.
We form the $n_0 \times n_0$ real orthogonal matrix $V$
by piling up all the row matrices $A_{JY}$ and $B_{JY}$ on top of each other.
Since $V$ is orthogonal,
\bs
\ba
\sum_{b=1}^{n_0} \left[ \left( A_{JY} \right)_{1b} \right]^2
= \sum_{b=1}^{n_0} \left[ \left( B_{JY} \right)_{1b} \right]^2
&=& 1,
\\
\sum_{b=1}^{n_0} \left( A_{JY} \right)_{1b} \left( B_{JY} \right)_{1b} &=& 0.
\ea
\es
Hence,
\bs
\ba
\sum_{b = 1}^{n_0 - 1} \sum_{b^\prime = b + 1}^{n_0}
\left[ \left( A_{JY} \right)_{1 b} \left( B_{JY} \right)_{1 b^\prime}
  - \left( B_{JY} \right)_{1 b} \left( A_{JY} \right)_{1 b^\prime} \right]^2 &=& 1,
\label{ofidfd}
\\
\sum_{b = 1}^{n_0} \left| \left( A_{JY} + i B_{JY} \right)_{1b} \right|^2 &=& 2.
\label{sum2}
\ea
\es

The masses of the gauge bosons are given in terms of the VEVs
of the scalar fields by
\bs
\label{eq:gaugemass}
\ba
\frac{m_Z^2}{2} &=& \frac{g^2}{c_W^2} \sum_{M_{JY} \in \AAA}
\left| v_{JY} \right|^2 Y^2,
\label{mz} \\
m_W^2 &=& g^2 \left[
  \sum_{M_{JY} \in \AAA \cap \RRR_1} \left| v_{JY} \right|^2
  \frac{\left( J + Y \right) \left( J - Y + 1 \right)}{2}
  \right. \no & & \left.
  + \sum_{M_{JY} \in \AAA \cap \SSS_1} \left| v_{JY} \right|^2
  \frac{\left( J + Y + 1 \right) \left( J - Y \right)}{2}
  \right]
\label{eq:mw1} \\ &=&
g^2 \left[
  \sum_{M_{JY} \in \AAA} \left| v_{JY} \right|^2
  \frac{\left( J + Y \right) \left( J - Y + 1 \right)}{2}
  + \sum_{M_{JY} \in \AAA} \left| v_{JY} \right|^2
  \frac{\left( J + Y + 1 \right) \left( J - Y \right)}{2}
  \right] \hspace*{5mm}
\label{eq:mw2} \\ &=&
g^2 \sum_{M_{JY} \in \AAA} \left| v_{JY} \right|^2 \left( J^2 - Y^2 + J \right).
\label{eq:mw3}
\ea
\es
In line~\eqref{eq:mw1} we have used firstly line~\eqref{u2} with $Q = 1$
and secondly line~\eqref{u3} with $Q = -1$.
In passing from line~\eqref{eq:mw1} to line~\eqref{eq:mw2}
we have used
\be
\label{setminus1}
M_{JY} \in \AAA \setminus \RRR_1 \Rightarrow J = -Y,
\quad \quad
M_{JY} \in \AAA \setminus \SSS_1 \Rightarrow J = Y.
\ee
One sees in Eqs.~\eqref{eq:gaugemass} that $m_Z^2 c_W^2 = m_W^2$
in general requires $J^2 + J = 3 Y^2$
for all the nonzero $v_{JY}$.\footnote{The relation~\eqref{relation}
may of course hold by accident,
even when $J \left( J + 1 \right) \neq 3 Y^2$ for some nonzero $v_{JY}$.}
This holds for the standard case of doublets with $J = Y = 1/2$,
and also for neutral singlets with $J = Y = 0$,
but it does not hold for most other choices of $J$ and $Y$.

\section{Prescription for \texorpdfstring{$S$}{S} and \texorpdfstring{$U$}{U}}
\label{sec:prescription}

\begin{figure}[t]
  \begin{center}
    \parbox{40mm}{
      \begin{fmfgraph*}(100,100)
        \fmfleft{i1}
        \fmfright{o1}
        \fmf{boson,label.side=left}{i1,w1}
        \fmf{boson,label.side=left}{w2,o1}
        \fmf{dashes,right,tension=.3}{w1,w2}
        \fmf{dashes,left,tension=.3}{w1,w2}
        \fmfv{lab=$f_\mu$,lab.dist=0.05w,lab.angle=0}{w1}
        \fmfv{lab=$f^\prime_\nu$,lab.dist=0.05w,lab.angle=180}{w2}
        \fmflabel{$V$}{i1}
        \fmflabel{$V^\prime$}{o1}
        \marrow{a}{down}{bot}{$q$}{i1,w1}{(1/4, 2/3)}
        \marrow{b}{down}{bot}{$q$}{w2,o1}{(1/3, 3/4)}
        \marrowii{c}{up}{top}{$p$}{w1,w2}{(1/4, 2/3)}
        \marrowii{d}{down}{bot}{$p^\prime$}{w1,w2}{(1/4, 2/3)}
    \end{fmfgraph*}}
  \end{center}
  \caption{One-loop vacuum-polarization diagram
    with two scalars as internal particles.}
  \label{fig:loopdiagram}
\end{figure}
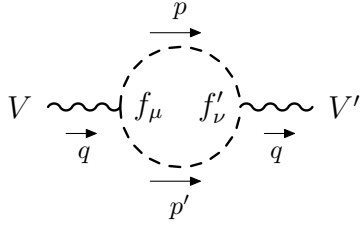

We consider the vacuum-polarization one-loop diagram
with two scalars as internal particles,
see Fig.~\ref{fig:loopdiagram}.
Let the Feynman rules for its vertices be
$f_\mu = K \left( p - p^\prime \right)_\mu$
and $f^\prime_\nu = K^\prime \left( p - p^\prime \right)_\nu$.
Then the contribution of that diagram to
\be
\label{uifgpo}
\left. \frac{\partial A_{V V^\prime} (q^2)}{\partial q^2} \right|_{q^2=0}
\ee
is
\be
\label{bjfgifo}
\left. \frac{\partial A_{V V^\prime} (q^2)}{\partial q^2} \right|_{q^2=0}
= - K K^\prime\, \mathrm{div} + \mathrm{finite\ terms},
\ee
where ``div'' is a divergent quantity defined in Eq.~\eqref{divdiv}
of appendix~\ref{sec:integrals}.
That quantity is independent both of $q^2$
and of the masses of the scalars in the diagram of Fig.~\ref{fig:loopdiagram}.

We want to compute the divergent contributions
to the quantitites~\eqref{uifgpo}
for $V V^\prime = AA$,
$AZ$,
$ZZ$,
and $WW$.
Those contributions originate solely in
diagrams like the one in Fig.~\ref{fig:loopdiagram}.
(Other types of diagrams give either vanishing or finite contributions.)
We therefore need the factors $K$ and $K^\prime$ for all the diagrams,
\textit{i.e.}\ we need the Lagrangian terms
that describe the interactions between one gauge boson and two scalars.
Those terms are derived from Eq.~\eqref{viw00e}
and may be found in appendix~\ref{sec:appendixLagrangian}.

Firstly consider $A_{AA} (q^2)$.
This is generated by loops with charged scalars $S_a^Q$ and $S_a^{-Q}$.
According to Eq.~\eqref{jvfigof} of appendix~\ref{sec:appendixLagrangian},
they generate
\be
\label{eq:AANPgeneral}
\left. \frac{\partial A_{AA} (q^2)}{\partial q^2} \right|_{q^2=0}
= e^2\, \mathrm{div}\, \sum_{Q > 0} Q^2 n_Q
+ \mathrm{finite\ terms}.
\ee

Secondly consider $A_{AZ} (q^2)$.
This is once again generated
by loops with charged scalars $S_a^Q$ and $S_a^{-Q}$.
According to Eqs.~\eqref{jvfigof} and~\eqref{bugf954}
of appendix~\ref{sec:appendixLagrangian},
they generate
\be
\left. \frac{\partial A_{AZ} (q^2)}{\partial q^2} \right|_{q^2=0}
=
\frac{e g}{c_W}\ \mathrm{div}\, \sum_{Q > 0}
\left( - c_W^2 Q^2 n_Q + \!\! \sum_{M_{JY} \in \RRR_Q} Q Y
- \!\! \sum_{M_{JY} \in \SSS_Q} Q Y \right)
+ \mathrm{finite\ terms},
\label{eq:AZNPgeneral}
\ee
where we have used Eqs.~\eqref{unitarity2} and~\eqref{unitarity3}.

Thirdly consider $A_{ZZ} (q^2)$.
This is generated
either by loops with charged scalars
$S_a^Q$ and $S_{a^\prime}^{-Q}$,
where $a^\prime$ may be different from $a$,
or by loops with neutral scalars
$S_b^0$ and $S_{b^\prime}^0$,
where $b^\prime$ may differ from $b$.
The relevant equations are Eqs.~\eqref{bugf954} and~\eqref{j0y77}
of appendix~\ref{sec:appendixLagrangian},
respectively.
They generate
\bs
\ba
\left. \frac{\partial A_{ZZ} (q^2)}{\partial q^2} \right|_{q^2=0}
&=& \frac{g^2}{c_W^2}\ \mathrm{div}\, \sum_{Q > 0}\
\sum_{a, a^\prime = 1}^{n_Q} \left| - c_W^2 Q\, \delta_{a a^\prime}
+ \sum_{M_{JY} \in \RRR_Q} Y \left( R^Q_{JY} \right)_{1a}
\left( R^Q_{JY} \right)_{1 a^\prime}^\ast \hspace*{7mm}
\right. \no & & \left.
- \sum_{M_{JY} \in \SSS_Q} Y \left( S^Q_{JY} \right)_{1a}
\left( S^Q_{JY} \right)_{1 a^\prime}^\ast \right|^2
\\ & &
+ \frac{g^2}{c_W^2}\ \mathrm{div}\, \sum_{b=1}^{n_0 - 1} \sum_{b^\prime = b+1}^{n_0}
\left\{ \sum_{M_{JY} \in \AAA} Y
\left[ \left( A_{JY} \right)_{1b} \left( B_{JY} \right)_{1 b^\prime}
  \right. \right. \no & & \left. \left.
  - \left( B_{JY} \right)_{1b} \left( A_{JY} \right)_{1 b^\prime} \right]
\vphantom{\sum_{Y = - J}^J
  Y \left( A_{JY} \right)_{1b} \left( B_{JY} \right)_{1 b^\prime}}
\right\}^2 + \mathrm{finite\ terms}.
\ea
\es
This leads to
\ba
\left. \frac{\partial A_{ZZ} (q^2)}{\partial q^2} \right|_{q^2=0}
&=& \frac{g^2}{c_W^2}\ \mathrm{div} \left[
\sum_{Q > 0} \left( c_W^4 Q^2 n_Q
- 2 c_W^2 Q \sum_{M_{JY} \in \RRR_Q} Y
+ 2 c_W^2 Q \sum_{M_{JY} \in \SSS_Q} Y
\right. \right. \hspace*{7mm} \no & & \left. \left.
 + \sum_{M_{JY} \in \RRR_Q} Y^2
 + \sum_{M_{JY} \in \SSS_Q} Y^2
\right)
+ \sum_{M_{JY} \in \AAA} Y^2 \right]
+ \mathrm{finite\ terms},
\label{eq:ZZNPgeneral}
\ea
where we have used Eqs.~\eqref{unitarity2}--\eqref{unitarity4}
and~\eqref{ofidfd}.

Finally consider $A_{WW} (q^2)$.
This is generated
either by loops with one charged scalar 
$S_a^Q$ and one neutral scalar $S_b^0$,
or by loops with two charged scalars
$S_a^{\pm Q}$ and $S_{a^\prime}^{\mp \left( Q + 1 \right)}$.
The relevant equations are Eqs.~\eqref{pri54} and~\eqref{pri55}
of appendix~\ref{sec:appendixLagrangian},
respectively.
They generate
\bs
\label{eq:WWNPgeneral}
\ba
\left. \frac{\partial A_{WW} (q^2)}{\partial q^2} \right|_{q^2=0}
&=& \frac{g^2}{2}\, \mathrm{div} \left\{
2 \sum_{M_{JY} \in \AAA} \left( J^2 - Y^2 + J\right) 
\right. \\ & &
+ \sum_{Q >0 } \left[
  \sum_{M_{JY} \in \RRR_Q}
  \left( J + Y - Q \right) \left( J - Y + Q + 1 \right)
  \right. \no & & \left. \left.
  + \sum_{M_{JY} \in \SSS_Q}
  \left( J + Y + Q + 1 \right) \left( J - Y - Q \right)
  \right]
\right\}
+ \mathrm{finite\ terms}, \hspace*{4mm}
\ea
\es
where we have used Eqs.~\eqref{unitarity2}--\eqref{unitarity4} and~\eqref{sum2},
and also
\be
\label{setminus2}
M_{JY} \in \RRR_Q \setminus \RRR_{Q+1} \Rightarrow J + Y - Q = 0,
\quad \quad
M_{JY} \in \SSS_Q \setminus \SSS_{Q+1} \Rightarrow J - Y - Q = 0.
\ee

It is now clear that the combinations relevant for the parameters $S$ and $U$
in Eqs.~\eqref{eq:S} and~\eqref{eq:U},
respectively,
namely
\bs
\label{eq:gammatheta}
\ba
\left. \frac{\partial A_{ZZ} (q^2)}{\partial q^2} \right|_{q^2=0}
-
\left. \frac{\partial A_{AA} (q^2)}{\partial q^2} \right|_{q^2=0}
& & \no
+ \frac{c_W^2 - s_W^2}{c_W s_W}\,
\left. \frac{\partial A_{AZ} (q^2)}{\partial q^2} \right|_{q^2=0}
&\equiv&
\frac{g^2}{c_W^2}\, \mathrm{div}\ \psi
+ \mathrm{finite\ terms},
\hspace*{5mm} \\
\left. \frac{\partial A_{WW} (q^2)}{\partial q^2} \right|_{q^2=0}
- c_W^2\,
\left. \frac{\partial A_{ZZ} (q^2)}{\partial q^2} \right|_{q^2=0}
& & \no
- s_W^2\,
\left. \frac{\partial A_{AA} (q^2)}{\partial q^2} \right|_{q^2=0}
+ 2 c_W s_W\, 
\left. \frac{\partial A_{AZ} (q^2)}{\partial q^2} \right|_{q^2=0}
&\equiv& \frac{g^2}{2}\, \mathrm{div}\ \theta + \mathrm{finite\ terms},
\ea
\es
are given by
\bs
\label{gammatheta2}
\ba
\psi &=& \sum_{M_{JY} \in \AAA} Y^2
  + \sum_{Q > 0} \left[
 \sum_{M_{JY} \in \RRR_Q} \left( Y^2 - QY \right)
+ \sum_{M_{JY} \in \SSS_Q} \left( Y^2 + QY \right)
\right],
\\
\theta &=&
2 \sum_{M_{JY} \in \AAA} \left( J^2 + J - 2 Y^2 \right)
+ \sum_{Q > 0} \left\{ \vphantom{\sum_{M_{JY} \in \SSS_Q}} - 2 Q^2 n_Q
\right. \no & &
+ \sum_{M_{JY} \in \RRR_Q}
\left[ \left( J + Y - Q \right) \left( J - Y + Q + 1 \right)
  - 2 Y^2 + 4 Q Y \right]
\no & & \left.
+ \sum_{M_{JY} \in \SSS_Q}
\left[ \left( J + Y + Q + 1 \right) \left( J - Y - Q \right)
  - 2 Y^2 - 4 Q Y \right] \right\}.
\ea
\es
It is proven in appendix~\ref{sec:appendixproof}
that $\psi$ and $\theta$ are both zero,
no matter what the set of $M_{JY}$
(and no matter what the VEVs of their neutral components are)
present in any particular $SU(2) \times U(1)$ electroweak model,
\textit{i.e.}\ either in the SM or in any extension thereof.

Since the quantities $\psi$ and $\theta$ vanish for the New Physics model and,
identically,
also for the Standard Model,
one might think that $S$ and $U$ would turn out finite.
However,
if we subtract the SM result for $\psi$ and $\theta$,
computed by using the usual SM Feynman rules for the Goldstone-boson vertices,
from the New Physics result,
which uses a \emph{different} set of Feynman rules for those vertices,
then we
get a gauge-dependent result for both $S$ and $U$.
This happens because the gauge-dependent parts of the NP and SM results
are different when Eq.~\eqref{relation} does not hold in the NP model.
Indeed,
as is shown in appendix~\ref{sec:appendixGoldstone},
some Goldstone-boson vertices \emph{depend on $\rho$},
hence they are different in the SM and in the NP model.
In Eqs.~\eqref{GBvertices} we display the relevant Goldstone-boson vertices
and their $\rho$-dependent value,
as derived in appendix~\ref{sec:appendixGoldstone};
we also display,
for each vertex and crossed over,
its value in the SM,
\textit{i.e.}\ when $\rho = 1$,
as given in Ref.~\cite{Branco1999}.
\vspace{5mm}
\bs
\label{GBvertices}
\ba
\parbox{35mm}{
    \begin{fmfgraph*}(75,75)
    \fmfleft{i1,i2}
    \fmfright{o1}
    \fmf{wiggly_arrow}{i1,w1}
    \fmf{photon}{i2,w1}
    \fmf{scalar}{o1,w1}
    \fmflabel{$W^{\mp}_\nu$}{i1}
    \fmflabel{$Z_\mu$}{i2}
    \fmflabel{$G^{\pm}$}{o1}
\end{fmfgraph*}}
&=& \ \xcancel{- i g s_W^2 m_Z g_{\mu \nu}}
\quad i g m_W \left( c_W - \frac{1}{\rho c_W} \right) g_{\mu \nu},
\label{eq:ZWGvertSM}\\[40pt]
\parbox{35mm}{
    \begin{fmfgraph*}(75,75)
    \fmfleft{i1,i2}
    \fmfright{o1}
    \fmf{scalar}{i1,w1}
    \fmf{scalar}{i2,w1}
    \fmf{photon}{w1,o1}
    \fmflabel{$G^-$}{i1}
    \fmflabel{$G^+$}{i2}
    \fmflabel{$Z_\mu$}{o1}
    \marrow{a}{left}{lft}{$p_-$}{i1,w1}{(1/4, 4/5)}
    \marrow{b}{left}{lft}{$p_+$}{i2,w1}{(1/4, 4/5)}
\end{fmfgraph*}}
&=& \ \xcancel{- i\ \frac{g \left( c_W^2 - s_W^2 \right)}{2 c_W}\,
  \left( p_- - p_+ \right)_\mu}
\quad i\, \frac{g}{2 c_W} \left( \frac{1}{\rho} - 2 c_W^2 \right)
\left( p_- - p_+ \right)_\mu,
\hspace*{9mm}
\label{eq:ZGGvertSM}
\\[40pt]
\parbox{35mm}{
    \begin{fmfgraph*}(75,75)
    \fmfleft{i1,i2}
    \fmfright{o1}
    \fmf{scalar}{i1,w1}
    \fmf{dashes}{i2,w1}
    \fmf{wiggly_arrow}{o1,w1}
    \fmflabel{$G^\mp$}{i1}
    \fmflabel{$G^0$}{i2}
    \fmflabel{$W_\mu^\pm$}{o1}
    \marrow{a}{left}{lft}{$q$}{i1,w1}{(1/4, 4/5)}
    \marrow{b}{left}{lft}{$p$}{i2,w1}{(1/4, 4/5)}
\end{fmfgraph*}}
&=& \ \xcancel{\frac{g}{2} \left ( q - p \right)_\mu} \quad
\frac{g}{2 \sqrt{\rho}}\left( q - p \right)_\mu.
\label{eq:WG0GvertSM}
\ea
\es
\vspace{5mm}

We thus have a problem.
We may obtain $\psi = \theta = 0$ in both the SM and its extension by using,
respectively,
the Feynman rules adequate for each of them,
with different values of $\rho$
(\textit{viz.}\ $\rho = 1$ in the SM and $\rho \neq 1$ in the extension);
but,
since the Feynman rules for the Goldstone-boson vertices
differ in the two cases,
we will then end up with a gauge-dependent result.
Or else we may use in both the SM and its extension the same Feynman rules,
as they are written in Eqs.~\eqref{GBvertices},
\textit{i.e.}\ with the same value of $\rho$,
and then we will get a gauge-independent result---but $\psi$ and $\theta$
will be non-zero in the SM and,
therefore,
$S$ and $U$,
respectively,
will diverge.

We propose to make $S$ and $U$ finite
by multiplying other SM Feynman rules---not the ones
in Eqs.~\eqref{GBvertices}---by factors that are equal to $1$ when $\rho = 1$.
This is possible because in the SM there is the Higgs particle $H$,
with mass $m_h$,
that participates in the computation of $\left. A_{ZZ} (q^2) \right|_\mathrm{SM}$
through a loop with $G^0$ and $H$,
and in the computation of $\left. A_{WW} (q^2) \right|_\mathrm{SM}$
through a loop with $G^+$ and $H$.
We thus propose to use,
in the SM,
the following Feynman rules for the vertices $Z G^0 H$,
$Z Z H$,
$W^\pm G^\mp H$,
and $W^+ W^- H$, respectively:
\vspace{5mm}
\bs
\label{newvertices}
\ba
\parbox{35mm}{
    \begin{fmfgraph*}(68,68)
    \fmfleft{i1,i2}
    \fmfright{o1}
    \fmf{dashes}{i1,w1}
    \fmf{dashes}{i2,w1}
    \fmf{photon}{w1,o1}
    \fmflabel{$H$}{i1}
    \fmflabel{$G^0$}{i2}
    \fmflabel{$Z^\mu$}{o1}
    \marrow{a}{left}{lft}{$p$}{i1,w1}{(1/4, 4/5)}
    \marrow{b}{left}{lft}{$q$}{i2,w1}{(1/4, 4/5)}
\end{fmfgraph*}}
&=& \frac{g}{2 c_W} \left( q - p \right)^\mu\, \sqrt{\aleph},
\\[40pt]
\parbox{35mm}{
    \begin{fmfgraph*}(68,68)
    \fmfleft{i1,i2}
    \fmfright{o1}
    \fmf{photon}{i1,w1}
    \fmf{photon}{i2,w1}
    \fmf{dashes}{w1,o1}
    \fmflabel{$Z^\nu$}{i1}
    \fmflabel{$Z^\mu$}{i2}
    \fmflabel{$H$}{o1}
\end{fmfgraph*}}
&=& i\ \frac{g m_Z}{c_W}\, g^{\mu \nu}\, \sqrt{\aleph},
\\[40pt]
\parbox{35mm}{
    \begin{fmfgraph*}(68,68)
    \fmfleft{i1,i2}
    \fmfright{o1}
    \fmf{dashes}{i1,w1}
    \fmf{scalar}{i2,w1}
    \fmf{wiggly_arrow}{o1,w1}
    \fmflabel{$H$}{i1}
    \fmflabel{$G^\mp$}{i2}
    \fmflabel{$W^{\pm \, \mu}$}{o1}
    \marrow{a}{left}{lft}{$p$}{i1,w1}{(1/4, 4/5)}
    \marrow{b}{left}{lft}{$q$}{i2,w1}{(1/4, 4/5)}
\end{fmfgraph*}}
&=& \mp i\ \frac{g}{2} \left( q - p \right)^\mu\, \sqrt{\beth},
\\[40pt]
\parbox{35mm}{
    \begin{fmfgraph*}(68,68)
    \fmfleft{i1,i2}
    \fmfright{o1}
    \fmf{wiggly_arrow}{i1,w1}
    \fmf{wiggly_arrow}{i2,w1}
    \fmf{dashes}{w1,o1}
    \fmflabel{$W^{- \, \nu}$}{i1}
    \fmflabel{$W^{+ \, \mu}$}{i2}
    \fmflabel{$H$}{o1}
\end{fmfgraph*}}
&=& i g m_W g^{\mu \nu}\, \sqrt{\beth}.
\ea
\es
\vspace{5mm}

The vertices $Z G^0 H$ and $Z Z H$ are multiplied
by the same factor $\sqrt{\aleph}$ due to gauge invariance.
The same happens with the vertices $W^\pm G^\mp H$ and $W^+ W^- H$,
which are both multiplied by $\sqrt{\beth}$.
Obviously,
in the true SM,
\textit{i.e.}\ when relation~\eqref{relation} holds,
both $\aleph$ and $\beth$
are~1.

In the computation of the SM contribution to $S$ and $U$,
we must consider six diagrams like the one in Fig.~\ref{fig:loopdiagram}:
\begin{enumerate}
\item A diagram with inner scalars $G^+$ and $H$
  contributing to $A_{WW} (q^2)$.
  \label{111}
\item A diagram with inner scalars $G^+$ and $G^0$
  contributing to $A_{WW} (q^2)$.
  \label{222}
\item A diagram with inner scalars $G^0$ and $H$
  contributing to $A_{ZZ} (q^2)$.
  \label{333}
\item A diagram with inner scalars $G^+$ and $G^-$
  contributing to $A_{ZZ} (q^2)$.
  \label{444}
\item A diagram with inner scalars $G^+$ and $G^-$
  contributing to $A_{AA} (q^2)$.
  \label{555}
\item A diagram with inner scalars $G^+$ and $G^-$
  contributing to $A_{AZ} (q^2)$.
  \label{666}
\end{enumerate}
A factor $\aleph$ multiplies the diagram~\ref{333}
and a factor $\beth$ multiplies the diagram~\ref{111}.

Using Eqs.~\eqref{eq:gammatheta}
and the results for the quantities~\eqref{uifgpo}
computed in the SM,
we get for the parameter $S$
\bs
\ba
\frac{\alpha}{4 s_W^2 c_W^2}\ S &=&
\frac{g^2}{c_W^2}\, \mathrm{div} \left[ \psi_\mathrm{NP} - \frac{\aleph}{4}
  - c_W^4 \left( 1 - \frac{m_Z^2}{2m_W^2} \right)^2
  + s_W^2 c_W^2
  \right. \\ & & \left.
  + c_W^2 \left( c_W^2 - s_W^2 \right)
  \left( 1 - \frac{m_Z^2}{2m_W^2} \right) \right]
+ \mathrm{finite\ terms},
\ea
\es
where the second to fifth terms inside the square brackets
originate in diagrams~\ref{333}--\ref{666},
respectively.
We have used Eq.~\eqref{eq:ZGGvertSM} in the third and fifth terms
inside the square brackets.
Since $\psi_\mathrm{NP} = 0$,
we
obtain a finite result for $S$ if
\be
\aleph = 4 \left[ - c_W^4 \left( 1 - \frac{m_Z^2}{2m_W^2} \right)^2
  + s_W^2 c_W^2
  + c_W^2 \left( c_W^2 - s_W^2 \right) \left( 1 - \frac{m_Z^2}{2m_W^2} \right)
  \right].
\ee
Thus,
we must choose
\be
\label{alephvalue}
\aleph = \frac{2}{\rho} - \frac{1}{\rho^2}.
\ee
Notice that $\aleph = 1$ when Eq.~\eqref{relation} holds,
\textit{viz.}\ when the extension of the SM only has
scalar doublets and/or singlets.

For the parameter $U$ we get
\bs
\ba
\frac{\alpha}{4 s_W^2}\, U &=&
\frac{g^2}{2}\, \mathrm{div} \left[ \theta_\mathrm{NP} - \frac{\beth}{2}
  - \frac{m_Z^2 c_W^2}{2m_W^2}
  + \frac{\aleph}{2}
  + 2 c_W^4 \left(1 - \frac{m_Z^2}{2 m_W^2} \right)^2
  \right. \\ & & \left.
  + 2 s_W^4
  + 4 s_W^2 c_W^2 \left( 1 - \frac{m_Z^2}{2 m_W^2} \right) \right]
+ \mathrm{finite\ terms},
\ea
\es
where the second to seventh terms inside the square brackets
originate in the diagrams~\ref{111}--\ref{666},
respectively.
We have used Eq.~\eqref{eq:WG0GvertSM} in the third term
and Eq.~\eqref{eq:ZGGvertSM} in the fifth and seventh terms
inside the square brackets.
Since $\theta_\mathrm{NP} = 0$,
we get a finite $U$ if
\be
\beth = - \frac{m_Z^2 c_W^2}{m_W^2} + \aleph
+ 4 c_W^4 \left(1 - \frac{m_Z^2}{2 m_W^2} \right)^2
+ 4 s_W^4 + 8 s_W^2 c_W^2 \left( 1 - \frac{m_Z^2}{2 m_W^2} \right).
\ee
Using Eq.~\eqref{alephvalue},
we obtain
\be
\label{bethvalue}
\beth = 4 - \frac{3}{\rho}.
\ee
Befittingly,
$\beth = 1$ when Eq.~\eqref{relation} holds,
\textit{viz.}\ in an extension of the SM
with only scalar doublets and singlets.

\section{The parameter \texorpdfstring{$S$}{S} in a model with scalar triplets}
\label{sec:triplets}

\subsection{The model}

In this section we consider an $SU(2) \times U(1)$ electroweak model
where the scalar sector comprises arbitrary numbers of
\begin{itemize}
\item $SU(2)$ doublets with weak hypercharge $1/2$
  \be
  \Phi_k=\begin{pmatrix}
  \phi_k^+  \\*[1mm] \phi_k^0
  \end{pmatrix} \quad (k=1, \ldots, n_d),
  \ee
\item $SU(2)$ triplets with weak hypercharge $1$
  \be
  \Xi_p=\begin{pmatrix}
  \xi_p^{++}  \\*[1mm] \xi_p^+   \\*[1mm] \xi_p^0
  \end{pmatrix} \quad (p=1, \ldots, n_{t_1}),
  \ee
\item real $SU(2)$ triplets with weak hypercharge $0$
  \be
  \Lambda_q=\begin{pmatrix}
  \lambda_q^+  \\*[1mm]
  \lambda_q^0 = {\lambda_q^0}^\ast  \\*[1mm]
  - {\lambda_q^+}^\ast
  \end{pmatrix} \quad (q=1, \ldots, n_{t_0}),
  \ee
\item $SU(2)$ singlets with weak hypercharge $1$
  \be
  \chi_j^+ \quad (j=1, \ldots, n_{s_1}),
  \ee
\item real $SU(2)$ singlets with weak hypercharge $0$
  \be
  \chi_l^0 = {\chi_l^0}^\ast \quad (l=1, \ldots, n_{s_0}),
  \ee
\item and $SU(2)$ singlets with weak hypercharge $2$
  \be
  \chi_r^{++} \quad (r=1, \ldots, n_{s_2}).
  \ee
\end{itemize}
There is then a total of
\begin{description}
\item $n_2=n_{t_1}+n_{s_2}$ complex scalar fields with electric charge $2$,
\item $n_1=n_d+n_{t_1}+n_{t_0}+n_{s_1}$
  complex scalar fields with electric charge $1$,
\item $n_0=2n_d+2n_{t_1}+n_{t_0}+n_{s_0}$ real scalar fields
  with electric charge $0$.
\end{description}

The neutral fields are allowed to have non-zero VEVs
\be
\left\langle 0 \left| \phi_k^0 \right| 0 \right\rangle
= \frac{v_k}{\sqrt{2}},
\quad \quad
\left\langle 0 \left| \xi_p^0 \right| 0 \right\rangle
= \frac{w_p}{\sqrt{2}},
\quad \quad
\left\langle 0 \left| \lambda_q^0 \right| 0 \right\rangle = x_q,
\quad \quad
\left\langle 0 \left| \chi_l^0 \right| 0 \right\rangle = u_l,
\ee
where the $v_k$ and the $w_p$ are in general complex
and the $x_q$ and the $u_l$ are real.
The masses of the gauge bosons $W^\pm$ and $Z^0$
are given in terms of the VEVs of the scalar fields as
\be
m_Z^2 = \frac{g^2}{4 c_W^2} \left( v^2 + 4 w^2 \right),
\qquad
m_W^2 = \frac{g^2}{4} \left( v^2 + 2 w^2 + 4 x^2 \right),
\label{mmwmmz}
\ee
where we have defined
\be
v = \sqrt{\sum_{k=1}^{n_d} \left| v_k \right|^2},
\qquad
w = \sqrt{\sum_{p=1}^{n_{t_1}} \left| w_p \right|^2},
\qquad
x = \sqrt{\sum_{q=1}^{n_{t_0}} x_q^2}.
\ee
Comparing the second Eq.~\eqref{mmwmmz} with Eq.~\eqref{eq:mw3},
note that there is an extra factor $1/2$ multiplying $x$.
This is because the field $\lambda_q^0$ is real,
hence it has a factor $1/2$ in its gauge-kinetic term.

We expand the neutral fields around their VEVs as 
\be
\label{eq:VEVs}
\phi_k^0 = \frac{v_k +  {\phi_k^0}^\prime}{\sqrt{2}},
\quad \quad
\xi_p^0 = \frac{w_p + {\xi_p^0}^\prime}{\sqrt{2}},
\quad \quad
\lambda_q^0 = x_q + {\lambda_q^0}^\prime,
\quad \quad
\chi_l^0 = u_l + {\chi_l^0}^\prime.
\ee
We define the mixing matrices of the scalar fields as
\be
\xi_p^{++} = \sum_{c=1}^{n_2} \left( T_2 \right)_{pc} S_c^{++},
\quad
\chi_r^{++} = \sum_{c=1}^{n_2} \left( T_4 \right)_{rc} S_c^{++},
\ee
\be
\phi_k^+ = \sum_{a=1}^{n_1} \left( U_1 \right)_{ka} S_a^+, \quad
\xi_p^+ = \sum_{a=1}^{n_1} \left( U_2 \right)_{pa} S_a^+, \quad
\lambda_q^+ = \sum_{a=1}^{n_1} \left( U_3 \right)_{qa} S_a^+, \quad
\chi_j^+ = \sum_{a=1}^{n_1} \left( U_4 \right)_{ja} S_a^+,
\ee
\be
{\phi_k^0}^\prime = \sum_{b=1}^{n_0} \left( V_1 \right)_{kb} S_b^0, \quad
{\xi_p^0}^\prime = \sum_{b=1}^{n_0} \left( V_2 \right)_{pb} S_b^0, \quad
{\lambda_q^0}^\prime = \sum_{b=1}^{n_0} \left( V_3 \right)_{qb} S_b^0, \quad
{\chi_l^0}^\prime = \sum_{b=1}^{n_0} \left( V_4 \right)_{lb} S_b^0,
\ee
where the dimensions of the matrices are
\bs
\ba
& &
T_2:\ n_{t_1} \times n_2, \quad \quad T_4:\ n_{s_2} \times n_2,
\\
& &
U_1:\ n_d \times n_1, \qquad
U_2:\ n_{t_1} \times n_1, \qquad
U_3:\ n_{t_0} \times n_1, \qquad
U_4:\ n_{s_1} \times n_1,
\\
& &
V_1:\ n_d \times n_0, \qquad V_2:\ n_{t_1} \times n_0, \qquad
V_3:\ n_{t_0} \times n_0, \qquad V_4:\ n_{s_0} \times n_0.
\ea
\es
The matrices $V_3$ and $V_4$ are real while the other ones are complex.
The matrix
\be
\tilde{T} = \begin{pmatrix} T_2 \\ T_4  \end{pmatrix}
\ee
is $n_2 \times n_2$ unitary;
it diagonalizes the Hermitian mass matrix of the scalars with charge $2$.
The matrix
\be
\label{eq:Util}
\tilde{U}=\begin{pmatrix} U_1 \\ U_2 \\ U_3 \\ U_4 \end{pmatrix}
\ee
is $n_1 \times n_1$ unitary;
it diagonalizes the Hermitian mass matrix of the scalars with charge $1$.
The matrix
\be
\label{eq:Vtil}
\tilde{V}=\begin{pmatrix}
\Re V_1  \\
\Im V_1  \\
\Re V_2  \\
\Im V_2  \\
V_3 \\
V_4
\end{pmatrix}
\ee
is $n_0 \times n_0$ real orthogonal;
it diagonalizes the symmetric mass matrix of the real components
of the neutral scalar fields.

One may obtain the form of the Goldstone bosons
by applying to the vacuum state the generators of the gauge group
that are spontaneously broken.
Using this method,
we obtain
\be
\label{eq:GoldstoneU}
\left( U_1 \right)_{k1} = \frac{v_k}{\sqrt{v^2 + 2 w^2 + 4 x^2}},\
\quad
\left( U_2 \right)_{p1} = \frac{\sqrt{2} w_p}{\sqrt{v^2 + 2 w^2 + 4 x^2}},
\quad
\left( U_3 \right)_{q1} = \frac{2 x_q}{\sqrt{v^2 + 2 w^2 + 4 x^2}},
\ee
\be
\label{eq:GoldstoneV}
\left( V_1 \right)_{k1} = \frac{i v_k}{\sqrt{v^2 + 4 w^2}},
\qquad
\left( V_2 \right)_{p1} = \frac{2 i w_p}{\sqrt{v^2 + 4w^2}},
\ee
while the first columns of $U_4$,
$V_3$,
and $V_4$ are identically zero.

\subsection{The formula for \texorpdfstring{$S$}{S}}

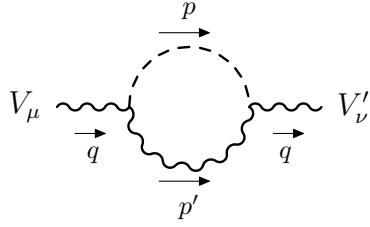
\begin{figure}[t]
  \begin{center}
    \parbox{40mm}{
      \begin{fmfgraph*}(100,100)
        \fmfleft{i1}
        \fmfright{o1}
        \fmf{boson,label.side=left}{i1,w1}
        \fmf{boson,label.side=left}{w2,o1}
        \fmf{boson,right,tension=.3}{w1,w2}
        \fmf{dashes,left,tension=.3}{w1,w2}
        \fmflabel{$V_\mu$}{i1}
        \fmflabel{$V^{\prime}_\nu$}{o1}
        \marrow{a}{down}{bot}{$q$}{i1,w1}{(1/4, 2/3)}
        \marrow{b}{down}{bot}{$q$}{w2,o1}{(1/3, 3/4)}
        \marrowii{c}{up}{top}{$p$}{w1,w2}{(1/4, 2/3)}
        \marrowii{d}{down}{bot}{$p^\prime$}{w1,w2}{(1/4, 2/3)}
    \end{fmfgraph*}}
  \end{center}
  \caption{One-loop vacuum-polarization diagram
    with one scalar and one gauge boson as internal particles.}
  \label{fig:loopdiagram2}
\end{figure}

There are two kinds of diagrams that produce the parameter $S$:
the ones like in Fig.~\ref{fig:loopdiagram}
and those like in Fig.~\ref{fig:loopdiagram2}.
Diagrams of the type of Fig.~\ref{fig:loopdiagram} produce
a function $K \left( I,\, J \right)$,
where $I$ and $J$ are the squared-masses of the two scalars in the loop.
Diagrams of the type of Fig.~\ref{fig:loopdiagram2} produce
a function $K \left( I,\, J \right) - 6 J\, \tilde K \left( I,\, J \right)$,
where $I$ is the squared-mass of the scalar
and $J$ is the squared-mass of the gauge boson in the loop.
The functions $K$ and $\tilde{K}$
are defined in Eqs.~\eqref{kk} of appendix~\ref{sec:integrals}.

Let $M_c$ denote the mass of the scalar $S_c^{++}$,
$m_a$ denote the mass of the scalar $S_a^+$,
and $\mu_b$ denote the mass of $S_b^0$.
Using the Feynman diagrams from appendix~\ref{sec:diagrams}
and the prescription described in section~\ref{sec:general},
we have obtained for the oblique parameter $S$ the following result:
\ba
S &=&
\frac{1}{12 \pi} \left\{
4\, \sum_{c,c^\prime =1}^{n_2}
\left| 2 s_W^2 \delta_{c c^\prime} - \left( T_2^\dagger T_2 \right)_{c c^\prime}
\right|^2 \left[ K \left( M_c^2,\, M_{c^\prime}^2 \right)
  + \frac{\ln{M_c^2} + \ln{M_{c^\prime}^2}}{2} \right]
\right. \no & &
+ 8\, \sum_{c=1}^{n_2} \left\{ \left( c_W^2 - s_W^2 \right)
\left[ 2 s_W^2 - \left( T_2^\dagger T_2 \right)_{c c} \right]
- 2 c_W^2 s_W^2 \right\} \ln{M_c^2}
\no & &
+ \sum_{a,a^\prime=2}^{n_1}
\left| 2 s_W^2 \delta_{a a^\prime} - \left( U_1^\dagger U_1 \right)_{a a^\prime}
- 2 \left( U_3^\dagger U_3 \right)_{a a^\prime}
\right|^2 \left[ K \left( m_a^2,\, m_{a^\prime}^2 \right)
  + \frac{\ln{m_a^2} + \ln{m_{a^\prime}^2}}{2} \right]
\no & &
+ 2\, \sum_{a=2}^{n_1} \left\{ \left( c_W^2 - s_W^2 \right)
\left[ 2 s_W^2 - \left( U_1^\dagger U_1 \right)_{a a}
  - 2 \left( U_3^\dagger U_3 \right)_{a a} \right]
- 2 c_W^2 s_W^2 \right\} \ln{m_a^2}
\no & &
+ \sum_{b=2}^{n_0 - 1} \sum_{b^\prime = b + 1}^{n_0}
\left[ \Im \left( V_1^\dagger V_1 \right)_{b b^\prime}
  + 2\, \Im \left( V_2^\dagger V_2 \right)_{b b^\prime}
  \right]^2 \left[ K \left( \mu_b^2,\, \mu_{b^\prime}^2 \right)
  + \frac{\ln{\mu_b^2} + \ln{\mu_{b^\prime}^2}}{2} \right]
\no & &
+ 2\, \sum_{a=2}^{n_1} \left| \left( U_2^\dagger U_2 \right)_{a 1}
- \left( U_3^\dagger U_3 \right)_{a 1} \right|^2
\left[ K \left( m_a^2,\, m_W^2 \right)
  + \frac{\ln{m_a^2} + \ln{m_W^2}}{2}
    \right. \no & & \left.
  \hspace*{60mm}  - 6 m_W^2\, \tilde{K} \left( m_a^2,\, m_W^2 \right) \right]
\no & &
+ \sum_{b=2}^{n_0} \left[ \Im \left( V_1^\dagger V_1 \right)_{1b}
  + 2\, \Im\left( V_2^\dagger V_2 \right)_{1b} \right]^2
\left[ K \left( \mu_b^2,\, m_Z^2 \right)
  + \frac{\ln{\mu_b^2} + \ln{m_Z^2}}{2}
    \right. \no & & \left.
  \hspace*{72mm} - 6 m_Z^2\, \Tilde{K} \left( \mu_b^2,\, m_Z^2 \right)
  \right]
\no & & \left.
- \left( \frac{2}{\rho} - \frac{1}{\rho^2} \right)
\left[ K \left( m_h^2,\, m_Z^2 \right)
  + \frac{\ln{m_h^2} + \ln{m_Z^2}}{2}
  - 6 m_Z^2\, \Tilde{K} \left( m_h^2,\, m_Z^2 \right)
  \right]
\right\}.
\label{eq:Sresulttriplets}
\ea
One may use the unitarity of the matrices $\tilde T$ and $\tilde U$
and the orthogonality of $\tilde V$ to write the following relations:
\ba
& & \sum_{a,a^\prime=2}^{n_1}
\left| 2 s_W^2 \delta_{a a^\prime} - \left( U_1^\dagger U_1 \right)_{a a^\prime}
- 2 \left( U_3^\dagger U_3 \right)_{a a^\prime}
\right|^2\, \ln m_a^2
\no &=&
\sum_{a=2}^{n_1} \left[ 4 s_W^4 + \left( 1 - 4 s_W^2 \right)
  \left( U_1^\dagger U_1 \right)_{a a} + 4 \left( 1 - 2 s_W^2 \right)
  \left( U_3^\dagger U_3 \right)_{a a} \right] \ln m_a^2
\no & &
- \sum_{a=2}^{n_1} \left| \left( U_2^\dagger U_2 \right)_{a 1}
- \left( U_3^\dagger U_3 \right)_{a 1} \right|^2 \ln m_a^2,
\label{yt1}
\ea
\ba
& & \frac{1}{2}\, \sum_{b=2}^{n_0 - 1} \sum_{b^\prime = b + 1}^{n_0}
\left[ \Im \left( V_1^\dagger V_1 \right)_{b b^\prime}
+ 2\, \Im \left( V_2^\dagger V_2 \right)_{b b^\prime}
\right]^2 \left( \ln \mu_b^2 + \ln \mu_{b^\prime}^2 \right)
\no &=&
\frac{1}{2}\, \sum_{b=2}^{n_0}
\left[\left( V_1^\dagger V_1 \right)_{b b}
  + 4 \left( V_2^\dagger V_2 \right)_{b b}\right] \ln \mu_b^2
\no & &
- \frac{1}{2}\, \sum_{b=2}^{n_0}
\left[ \Im \left( V_1^\dagger V_1 \right)_{1b}
  + 2\, \Im\left( V_2^\dagger V_2 \right)_{1b} \right]^2 \ln \mu_b^2,
\label{yt2}
\ea
\be
\sum_{b=2}^{n_0} \left[ \Im \left( V_1^\dagger V_1 \right)_{1b}
  + 2\, \Im\left( V_2^\dagger V_2 \right)_{1b} \right]^2
= \left( V_1^\dagger V_1 \right)_{1 1} + 4 \left( V_2^\dagger V_2 \right)_{1 1},
\label{yt3}
\ee
\be
\sum_{a=2}^{n_1} \left| \left( U_2^\dagger U_2 \right)_{a 1}
- \left( U_3^\dagger U_3 \right)_{a 1} \right|^2
= \left( U_2^\dagger U_2 \right)_{1 1}
+ \left( U_3^\dagger U_3 \right)_{1 1}
- \left( 1 - \frac{1}{\rho} \right)^2,
\label{yt4}
\ee
\be
\sum_{c,c^\prime =1}^{n_2}
\left| 2 s_W^2 \delta_{c c^\prime}
- \left( T_2^\dagger T_2 \right)_{c c^\prime} \right|^2\, \ln M_c^2
= \sum_{c =1}^{n_2} \left[ 4 s_W^4 + \left( 1 - 4 s_W^2 \right)
  \left( T_2^\dagger T_2 \right)_{c c} \right] \ln M_c^2.
\label{yt5}
\ee
We use Eqs.~\eqref{yt1}--\eqref{yt5}
to simplify Eq.~\eqref{eq:Sresulttriplets},
obtaining
\bs
\label{eq:Sresulttriplets2}
\ba
S &=&
\frac{1}{12 \pi} \left\{
4\, \sum_{c,c^\prime =1}^{n_2}
\left| 2 s_W^2 \delta_{c c^\prime} - \left( T_2^\dagger T_2 \right)_{c c^\prime}
\right|^2 K \left( M_c^2,\, M_{c^\prime}^2 \right)
\right. \\ & &
+ \sum_{a,a^\prime=2}^{n_1}
\left| 2 s_W^2 \delta_{a a^\prime} - \left( U_1^\dagger U_1 \right)_{a a^\prime}
- 2 \left( U_3^\dagger U_3 \right)_{a a^\prime}
\right|^2 K \left( m_a^2,\, m_{a^\prime}^2 \right)
\\ & &
+ \sum_{b=2}^{n_0 - 1} \sum_{b^\prime = b + 1}^{n_0}
\left[ \Im \left( V_1^\dagger V_1 \right)_{b b^\prime}
  + 2\, \Im \left( V_2^\dagger V_2 \right)_{b b^\prime}
  \right]^2 K \left( \mu_b^2,\, \mu_{b^\prime}^2 \right)
\\ & &
+ 2\, \sum_{a=2}^{n_1} \left| \left( U_2^\dagger U_2 \right)_{a 1}
- \left( U_3^\dagger U_3 \right)_{a 1} \right|^2
\left[ K \left( m_a^2,\, m_W^2 \right)
  - 6 m_W^2\, \tilde{K} \left( m_a^2,\, m_W^2 \right) \right]
\\ & &
+ \sum_{b=2}^{n_0} \left[ \Im \left( V_1^\dagger V_1 \right)_{1b}
  + 2\, \Im\left( V_2^\dagger V_2 \right)_{1b} \right]^2
\left[ K \left( \mu_b^2,\, m_Z^2 \right)
  - 6 m_Z^2\, \Tilde{K} \left( \mu_b^2,\, m_Z^2 \right) \right]
\\ & &
- \left( \frac{2}{\rho} - \frac{1}{\rho^2} \right)
\left[ K \left( m_h^2,\, m_Z^2 \right)
  - 6 m_Z^2\, \Tilde{K} \left( m_h^2,\, m_Z^2 \right) \right]
\\ & &
- 4\, \sum_{c=1}^{n_2} \left( T_2^\dagger T_2 \right)_{c c} \ln{M_c^2}
- \sum_{a=2}^{n_1} \left( U_1^\dagger U_1 \right)_{a a} \ln{m_a^2}
\label{uy1} \\ & &
+ \sum_{b=2}^{n_0}
\left[ \left( V_1^\dagger V_1 \right)_{b b}
  + 4 \left( V_2^\dagger V_2 \right)_{b b} \right]
\frac{\ln{\mu_b^2}}{2}
\label{uy2} \\ & &
+ \left[ \left( V_1^\dagger V_1 \right)_{11}
  + 4 \left( V_2^\dagger V_2 \right)_{11} \right]
\frac{\ln{m_Z^2}}{2}
\label{uy3} \\ & &
+ \left[\left( U_2^\dagger U_2 \right)_{1 1}
  + \left( U_3^\dagger U_3 \right)_{1 1}
  - \left( 1 - \frac{1}{\rho} \right)^2 \right]
\ln m_{W}^2
\label{uy4} \\ & & \left.
- \left( \frac{2}{\rho} - \frac{1}{\rho^2} \right)
\frac{\ln{m_h^2} + \ln{m_Z^2}}{2} \right\}.
\label{uy5}
\ea
\es
The coefficients of the logarithms
in lines~\eqref{uy1}--\eqref{uy5} add up tp zero as they should:
\bs
\ba
& &
- 4\, \sum_{c=1}^{n_2} \left( T_2^\dagger T_2 \right)_{c c}
- \sum_{a=2}^{n_1} \left( U_1^\dagger U_1 \right)_{a a}
+ \frac{1}{2}\ \sum_{b=1}^{n_0}
\left[ \left( V_1^\dagger V_1 \right)_{bb}
  + 4 \left( V_2^\dagger V_2 \right)_{bb} \right]
\\ & &
+ \left( U_2^\dagger U_2 \right)_{11}
+ \left( U_3^\dagger U_3 \right)_{11}
- \left( 1 - \frac{1}{\rho} \right)^2
- \left( \frac{2}{\rho} - \frac{1}{\rho^2} \right)
\\ &=&
- 4\, \mathrm{tr} \left( T_2 T_2^\dagger \right)
- \mathrm{tr} \left( U_1 U_1^\dagger \right)
+  \left( U_1^\dagger U_1 \right)_{11}
+ \frac{1}{2}\, \mathrm{tr} \left( V_1 V_1^\dagger \right)
+ 2\, \mathrm{tr} \left( V_2 V_2^\dagger \right)
\no & &
+ \left( U_2^\dagger U_2 \right)_{11}
+ \left( U_3^\dagger U_3 \right)_{11}
- 1
\\ &=&
- 4 n_{t_1}
- n_d
+  \left( U_1^\dagger U_1 \right)_{11}
+ n_d
+ 4 n_{t_1}
+ \left( U_2^\dagger U_2 \right)_{11}
+ \left( U_3^\dagger U_3 \right)_{11}
- 1
\\ &=&
\left( U_1^\dagger U_1 \right)_{11}
+ \left( U_2^\dagger U_2 \right)_{11}
+ \left( U_3^\dagger U_3 \right)_{11} - 1
\\ &=& 0,
\ea
\es
where in the last step we have used Eqs.~\eqref{eq:GoldstoneU}.

Equation~\eqref{eq:Sresulttriplets2} for $S$
generalizes the result given in ref.~\cite{Grimus20082}
for the case without any scalar $SU(2)$ triplets
and without charge-$2$ singlets.
In that case the matrices $T_2$,
$T_4$,
$U_2$,
$U_3$,
$V_2$,
and $V_3$ and zero.
Ref.~\cite{Grimus20082} then gave
\bs
\label{oldS}
\ba
S &=& \frac{1}{24 \pi} \left\{
\sum_{a=2}^{n_1} \left[ 2 s_W^2 - \left( U_1^\dagger U_1 \right)_{aa} \right]^2
G \left( m_a^2,\, m_a^2,\, m_Z^2 \right)
\right. \\ & &
+ 2\, \sum_{a=2}^{n_1-1} \sum_{a^\prime=a+1}^{n_1}
\left| \left( U_1^\dagger U_1 \right)_{a a^\prime} \right|^2
G \left( m_a^2,\, m_{a^\prime}^2,\, m_Z^2 \right)
\\ & &
+ \sum_{b=2}^{n_0-1} \sum_{b^\prime=b+1}^{n_0}
\left[ \mathrm{Im} \left( V_1^\dagger V_1 \right)_{b b^\prime} \right]^2
G \left( \mu_b^2,\, \mu_{b^\prime}^2,\, m_Z^2 \right)
\\ & &
+ \sum_{b=2}^{n_0}
\left[ \mathrm{Im} \left( V_1^\dagger V_1 \right)_{1 b} \right]^2
\hat G \left( \mu_b^2,\, m_Z^2 \right)
- \hat G \left( m_h^2,\, m_Z^2 \right)
\\ & & \left.
- \ln{m_h^2}
- 2\, \sum_{a=2}^{n_1} \left( U_1^\dagger U_1 \right)_{aa} \ln{m_a^2}
+ \sum_{b=2}^{n_0} \left( V_1^\dagger V_1 \right)_{bb} \ln{\mu_b^2}
\right\}.
\ea
\es
It must be stressed that ref.~\cite{Grimus20082}
used a different definition of $S$~\cite{maksymyk},
wherein
\be
\left. \frac{\partial \, \delta A_{Z Z}(q^2)}{\partial q^2} \right|_{q^2=0}
\ee
in Eq.~\eqref{eq:S} was substituted by
\be
\frac{\delta A_{Z Z} \left( m_Z^2 \right)
  - \delta A_{Z Z} \left( 0 \right)}{m_Z^2}.
\ee
This caused the appearance of the functions $G \left( I,\, J,\, m_Z^2 \right)$
and $\hat G \left( J,\, m_Z^2 \right)$ in Eq.~\eqref{oldS}.
In order to make the connection with the present work,
one must use
\bs
\ba
G \left( I,\, J,\, m_Z^2 \right)
&=& 2\, K \left( I,\, J \right) + O \left( m_Z^2 \right),
\\
\hat G \left( J,\, m_Z^2 \right)
&=& 2 \left[ K \left( J,\, m_Z^2 \right)
  - 6 m_Z^2\, \tilde K \left( J,\, m_Z^2 \right) \right]
+ O \left( m_Z^2 \right).
\ea
\es

\section{Conclusions}
\label{sec:conclusions}

In this paper we have discovered a problem
that arises in the computation of the oblique parameters $S$ and $U$
whenever the New-Physics model at hand does not obey relation~\eqref{relation},
because of scalars in multiplets of $SU(2)$ larger than doublets having VEVs.
Namely,
the Feynman rules for some Goldstone-boson vertices depend on $\rho$---see
Eqs.~\eqref{GBvertices}---and this causes a mismatch
between the New-Physics model,
wherein $\rho \neq 1$,
and the Standard Model,
wherein $\rho = 1$.
Depending on how one (mis)handles this mismatch,
$S$ and $U$ may turn out either gauge-dependent or divergent.
We have proceeded by suggesting a solution for this problem
through a redefinition of the Standard-Model Feynman rules
for some vertices containing the Higgs boson $H$.
Namely,
we have inserted by hand into those Feynman rules
factors $\sqrt{\aleph}$ and $\sqrt{\beth}$---see
Eqs.~\eqref{newvertices}---that are equal to $1$ when relation~\eqref{relation} holds.
One easily finds out that,
if $\aleph$ and $\beth$ take the values in Eqs.~\eqref{alephvalue}
and~\eqref{bethvalue},
respectively,
then the problem is solved.

We have utilized our insight to explicitly compute the value of $S$
in a New-Physics model containing arbitrary numbers of triplets of $SU(2)$
(and also extra doublets and singlets)
with weak hypercharges either $0$ or $1$.

In the future,
it may be interesting to study whether the changes to the Feynman rules
that we have suggested may somehow be justified,
whether they cure the problems in the computation of $S$ and $U$
also at the two-loop level,
and whether they may cure other possible problems elsewhere.
We leave these questions for consideration by other authors.

\vspace*{5mm}

\paragraph{Acknowledgements:}
This work was supported by the Portuguese Foundation for Science and Technology
through projects CERN/FIS-PAR/0004/2019,
CERN/FIS-PAR/0008/2019,
PTDC/FIS-PAR/29436/2017,
UIDB/00777/2020,
and UIDP/00777/2020.
F.A.~acknowledges a fellowship from Project CERN/FIS-PAR/0008/2019.

\newpage

\appendix

\section{Feynman integrals}
\label{sec:integrals}

We define the function
\be
D \left( Q, I, J, x \right) =
Q x \left( x - 1 \right) + I \left( 1 - x \right) + J x.
\ee
We have used the following results:
\bs
\label{ints}
\ba
\int_0^1 \mathrm{d}x
\left. \frac{\partial}{\partial Q} \left\{ \frac{4}{d}\ \mu^{4-d}
\int \frac{\mathrm{d}^d k}{\left( 2 \pi \right)^d}\,
\frac{k^2}{\left[ k^2 - D \left( Q, I, J, x \right) + i \epsilon \right]^2}
\right\} \right|_{Q=0}
&=& i \left[ \text{div} + \frac{K \left( I, J \right)}{48 \pi^2}
  \right. \no & & \left.
  + \frac{\ln I +  \ln J}{96 \pi^2} \right],
\\
\int_0^1 \mathrm{d}x \left. \frac{\partial}{\partial Q}
\left\{ \mu^{4-d} \int \frac{\mathrm{d}^d k}{\left( 2 \pi \right)^d}\
\frac{1}{\left[ k^2
    - D \left( Q, I, J, x \right)
    + i \epsilon\right]^2}
\right\} \right|_{Q=0}
&=& \frac{i}{32 \pi^2}\ \Tilde{K} \left( I, J \right),
\ea
\es
where
\bs
\label{kk}
\ba
K \left( I, J \right) &=& \left\{ \begin{array}{lcl}
  {\displaystyle - \frac{5}{6} + \frac{2 I J}{\left( I - J \right)^2}
    + \frac{I^3 + J^3 - 3 I J \left( I + J \right)}{2 \left( I - J \right)^3}\,
    \ln{\frac{I}{J}}} &\Leftarrow& I \neq J,
  \\
  0 &\Leftarrow& I = J,
\end{array} \right.
\\
\Tilde{K} \left( I, J \right) &=& \left\{ \begin{array}{lcl}
  {\displaystyle \frac{I + J}{\left( I - J \right)^2}
    - \frac{2 I J}{\left( I - J \right)^3}\,
    \ln{\frac{I}{J}}} &\Leftarrow& I \neq J,
  \\*[5mm]
    {\displaystyle \frac{1}{3I}} & \Leftarrow& I = J,
\end{array} \right.
\ea
\es
and
\be
\mathrm{div} = \frac{- 1}{48 \pi^2} \left[ \frac{2}{4 - d}
  - \gamma + \ln{\left( 4 \pi \mu^2 \right)} \right].
\label{divdiv}
\ee
In Eqs.~\eqref{ints} and~\eqref{divdiv},
$\mu$ is an arbitrary mass scale.
In Eq.~\eqref{divdiv},
$\gamma$ is Euler's constant.
The quantity named ``$\mathrm{div}$'' diverges in the limit $d \to 4$.

\section{Interaction terms of one gauge boson with two scalars}
\label{sec:appendixLagrangian}

In this appendix we derive some consequences of Eq.~\eqref{viw00e}.

The interaction terms of the photon with the charge-$Q$ scalars
($Q > 0$)
are given by
\bs
\label{jvfigof}
\ba
\mathcal{L}_{A S^Q S^{-Q}} &=& i e Q A^\mu \left\{
\sum_{M_{JY} \in \RRR_Q}
\left[ M_{JY}^Q\, \partial_\mu \left( M_{JY}^Q \right)^\ast -
  \left( M_{JY}^Q \right)^\ast \partial_\mu M_{JY}^Q \right]
\right. \\ & & \left.
+ \sum_{M_{JY} \in \SSS_Q}
\left[ \left( M_{JY}^{-Q} \right)^\ast \partial_\mu M_{JY}^{-Q}
  - M_{JY}^{-Q}\, \partial_\mu \left( M_{JY}^{-Q} \right)^\ast \right]
\right\}
\\ &=& i e Q A^\mu \sum_{a,a^\prime=1}^{n_Q}
\left( S_a^Q\, \partial_\mu S_{a^\prime}^{-Q}
- S_{a^\prime}^{-Q}\, \partial_\mu S_a^Q \right)
\\ & &
\times \left[
  \sum_{M_{JY} \in \RRR_Q} \left( R_{JY}^Q \right)_{1 a}
  \left( R_{JY}^Q \right)^\ast_{1 a^\prime}
  +
  \sum_{M_{JY} \in \SSS_Q} \left( S_{JY}^Q \right)_{1 a}
  \left( S_{JY}^Q \right)^\ast_{1 a^\prime}
  \right]
\\ &=& i e Q A^\mu \sum_{a=1}^{n_Q} \left( S_a^Q\, \partial_\mu S_{a}^{-Q}
- S_{a}^{-Q}\, \partial_\mu S_a^Q \right),
\ea
\es
where we have used Eq.~\eqref{unitarity1}.

The interaction terms of the $Z$ with two charged scalars are given by
\bs
\label{bugf954}
\ba
\mathcal{L}_{Z S^Q S^{-Q}} &=& i\, \frac{g}{c_W}\, Z^\mu \left\{
\sum_{M_{JY} \in \RRR_Q}
\left( Y - Q c_W^2 \right)
\left[ M_{JY}^Q\, \partial_\mu \left( M_{JY}^Q \right)^\ast
  - \left( M_{JY}^Q \right)^\ast \partial_\mu M_{JY}^Q \right]
\right.  \hspace*{7mm} \\ & & \left.
- \sum_{M_{JY} \in \SSS_Q}
\left( Y + Q c_W^2 \right)
\left[ \left( M_{JY}^{-Q} \right)^\ast \partial_\mu M_{JY}^{-Q}
  - M_{JY}^{-Q}\, \partial_\mu \left( M_{JY}^{-Q} \right)^\ast \right]
\right\}
\\ &=&
- i g c_W Q Z^\mu \sum_{a=1}^{n_Q} \left( S_a^Q\, \partial_\mu S_{a}^{-Q}
- S_{a}^{-Q}\, \partial_\mu S_a^Q \right)
\\ & &
+ i\, \frac{g}{c_W}\, Z^\mu
\sum_{a,a^\prime=1}^{n_Q} \left( S_a^Q\, \partial_\mu S_{a^\prime}^{-Q}
- S_{a^\prime}^{-Q}\, \partial_\mu S_a^Q \right)
\\ & &
\times \left[
  \sum_{M_{JY} \in \RRR_Q} Y \left( R_{JY}^Q \right)_{1 a}
  \left( R_{JY}^Q \right)^\ast_{1 a^\prime}
-
\sum_{M_{JY} \in \SSS_Q} Y \left( S_{JY}^Q \right)_{1 a}
\left( S_{JY}^Q \right)^\ast_{1 a^\prime}
\right].
\ea
\es

The interaction terms of the $Z$ with two neutral scalars are given by
\bs
\label{j0y77}
\ba
\mathcal{L}_{Z S^0 S^0} &=& i\, \frac{g}{c_W}\,
Z^\mu \!\! \sum_{M_{JY} \in \AAA} Y
\left[ M_{JY}^0\, \partial_\mu \left( M_{JY}^0 \right)^\ast
  - \left( M_{JY}^0 \right)^\ast \partial_\mu M_{JY}^0 \right]
\\
&=& i\, \frac{g}{2c_W}\, Z^\mu \sum_{b,b^\prime=1}^{n_0}
\left( S_b^0\, \partial_\mu S_{b^\prime}^0
- S_{b^\prime}^0\, \partial_\mu S_{b}^0 \right)
\\ & &
\times \sum_{M_{JY} \in \AAA}
Y \left[ \left( A_{JY} \right)_{1b} + i \left( B_{JY} \right)_{1 b}\right]
\left[ \left( A_{JY} \right)_{1b^\prime}
  - i \left( B_{JY} \right)_{1 b^\prime} \right]
\\
&=&
\frac{g}{c_W}\, Z^\mu \sum_{b=1}^{n_0 - 1} \sum_{b^\prime = b+1}^{n_0}
\left( S_b^0\, \partial_\mu S_{b^\prime}^0
- S_{b^\prime}^0\, \partial_\mu S_{b}^0 \right)
\\ & &
\times \sum_{M_{JY} \in \AAA} Y \left[
  \left( A_{JY} \right)_{1b} \left( B_{JY} \right)_{1 b^\prime}
  - \left( A_{JY} \right)_{1b^\prime} \left( B_{JY} \right)_{1 b}
  \right].
\ea
\es

The interaction terms of the $W$ with
one singly-charged scalar
and one neutral scalar are given by
\bs
\label{pri54}
\ba
\mathcal{L}_{W^\pm S^\mp S^0} &=&
i g \sum_{M_{JY} \in \AAA \cap \SSS_1}
\sqrt{\frac{\left( J + Y + 1 \right) \left( J - Y \right)}{2}}
\\*[1mm] & & \times
\left\{ W^{\mu +} \left[
  \left( M_{JY}^{0} \right)^\ast\, \partial_\mu M_{JY}^{-1}
  - M_{JY}^{-1}\, \partial_\mu \left( M_{JY}^0 \right)^\ast \right]
  \right. \\*[0.5mm] & & \left.
  + W^{\mu -} \left[ \left( M_{JY}^{-1} \right)^\ast\, \partial_\mu M_{JY}^0
    - M_{JY}^0\, \partial_\mu \left( M_{JY}^{-1} \right)^\ast \right]
  \right\}
\\*[1mm] & &
+ i g \sum_{M_{JY} \in \AAA \cap \RRR_1}
\sqrt{\frac{\left( J + Y \right) \left( J - Y + 1 \right)}{2}}
\\*[1mm] & & \times
\left\{ W^{\mu +} \left[
  \left( M_{JY}^1 \right)^\ast \partial_\mu M_{JY}^0
  - M_{JY}^0\, \partial_\mu \left( M_{JY}^1 \right)^\ast
  \right]
  \right. \\*[0.5mm] & & \left.
  + W^{\mu -} \left[ \left( M_{JY}^0 \right)^\ast \partial_\mu M_{JY}^1
    - M_{JY}^1\, \partial_\mu \left( M_{JY}^0 \right)^\ast \right]
  \right\}
\\ &=&
i\, \frac{g}{2}\, \sum_{M_{JY} \in \AAA}\,
\sum_{a=1}^{n_1}\, \sum_{b=1}^{n_0}\,
\sqrt{\left( J + Y + 1 \right) \left( J - Y \right)}
\\*[1mm] & & \times
\left[ W^{\mu +} \left( A_{JY} - i B_{JY} \right)_{1b}
  \left( S_{JY}^1 \right)^\ast_{1a}
  \left( S_b^0\, \partial_\mu S_a^- - S_a^-\, \partial_\mu S_b^0 \right)
  \right. \\*[0.5mm] & & \left.
  + W^{\mu -} \left( A_{JY} + i B_{JY} \right)_{1b} \left( S_{JY}^1 \right)_{1a}
  \left( S_a^+\, \partial_\mu S_b^0 - S_b^0\, \partial_\mu S_a^+ \right)
  \right]
\\ & & + i\, \frac{g}{2}\, \sum_{M_{JY} \in \AAA}\,
\sum_{a=1}^{n_1}\, \sum_{b=1}^{n_0}\,
\sqrt{\left( J + Y \right) \left( J - Y + 1 \right)}
\\*[1mm] & & \times
\left[ W^{\mu +} \left( A_{JY} + i B_{JY} \right)_{1b}
  \left( R_{JY}^1 \right)^\ast_{1a}
  \left( S_a^-\, \partial_\mu S_b^0  - S_b^0\, \partial_\mu S_a^- \right)
  \right. \\*[0.5mm] & & \left.
  + W^{\mu -} \left( A_{JY} - i B_{JY} \right)_{1b} \left( R_{JY}^1 \right)_{1a}
  \left (S_b^0\, \partial_\mu S_a^+  - S_a^+\, \partial_\mu S_b^0 \right)
  \right],
\ea
\es
where we have used Eqs.~\eqref{setminus1}.

The interaction terms of the $W$ with two charged scalars are given by
\bs
\label{pri55}
\ba
\mathcal{L} &=& \cdots +
i\, \frac{g}{\sqrt{2}}\, \sum_{a=1}^{n_Q}\, \sum_{a^\prime = 1}^{n_{Q+1}}\, \left\{
W_\mu^+\, \left( S_a^Q\, \partial^\mu S_{a^\prime}^{-Q-1}
- S_{a^\prime}^{-Q-1}\, \partial^\mu S_a^Q \right)
\right. \\ & & \times \left[
  \sum_{M_{JY} \in \SSS_Q \cap \SSS_{Q+1}} \!
  \sqrt{\left( J + Y + Q + 1 \right) \left( J - Y - Q \right)}\,
  \left( S_{JY}^Q \right)_{1a} \left( S_{JY}^{Q+1} \right)^\ast_{1 a^\prime}
  \right. \\ & & \left.
  - \!\! \sum_{M_{JY} \in \RRR_Q \cap \RRR_{Q+1}} \!
  \sqrt{\left( J + Y - Q \right) \left( J - Y + Q + 1 \right)}\,
  \left( R_{JY}^Q \right)_{1a} \left( R_{JY}^{Q+1} \right)_{1a^\prime}^\ast
  \right]
\\ & &
+ W_\mu^-
\left( S_{a^\prime}^{Q+1}\, \partial_\mu S_a^{-Q}
- S_a^{-Q}\, \partial_\mu S_{a^\prime}^{Q+1} \right)
\\ & & \times \left[
\sum_{M_{JY} \in \SSS_Q \cap \SSS_{Q+1}} \!
  \sqrt{\left( J + Y + Q + 1 \right) \left( J - Y - Q \right)}\,
  \left( S_{JY}^Q \right)^\ast_{1a} \left( S_{JY}^{Q+1} \right)_{1 a^\prime}
  \right. \\ & & \left. \left.
  - \!\! \sum_{M_{JY} \in \RRR_Q \cap \RRR_{Q+1}} \!
  \sqrt{\left( J + Y - Q \right) \left( J - Y + Q + 1 \right)}\,
  \left( R_{JY}^Q \right)_{1a}^\ast \left( R_{JY}^{Q+1} \right)_{1a^\prime}
  \right] \right\}.
\ea
\es

\section{Proof of \texorpdfstring{$\psi = \theta = 0$}{psi = theta = 0}}
\label{sec:appendixproof}

We show in this appendix that the quantities $\psi$ and $\theta$
in Eqs.~\eqref{gammatheta2} are equal to $0$.
We shall use the sums
\bs
\ba
1 + 2 + 3 + \cdots + n &=& \frac{n \left( n + 1 \right)}{2},
\\
1 + 4 + 9 + \cdots + n^2 &=&
\frac{n \left( n + 1 \right) \left( 2 n + 1 \right)}{6},
\ea
\es
which are valid for any positive integer $n$.

We shall separately consider the following types of multiplets $M_{JY}$:
\begin{enumerate}
\item $Y < - J$. \label{type1}
  A multiplet of this type belongs neither to $\AAA$ nor to $\RRR_Q$
  for any $Q > 0$.
  It belongs to $\SSS_Q$ for $Q = - Y - J, \ldots, - Y + J$.
\item $Y = - J$. \label{type2}
  A multiplet of this type belongs to $\AAA$.
  It also belongs to $\SSS_Q$ for $Q = 1, 2, \ldots, - 2 Y$.
  It does not belong to $\RRR_Q$ for any value of $Q > 0$.
\item $- J < Y < J$. \label{type3}
  A multiplet of this type belongs to $\AAA$,
  to $\RRR_Q$ for $Q = 1, \ldots, J + Y$,
  and to $\SSS_Q$ for $Q = 1, \ldots, J - Y$.
\item $Y = J$. \label{type4}
  A multiplet of this type belongs to $\AAA$
  and also to $\RRR_Q$ for $Q = 1, 2, \ldots, 2 Y$.
  It does not belong to $\SSS_Q$ for any value of $Q > 0$.
\item $Y > J$. \label{type5}
  A multiplet of this type belongs neither to $\AAA$
  nor to $\SSS_Q$ for any $Q > 0$.
  It belongs to $\RRR_Q$ for $Q = Y - J, \ldots, Y + J$.
\end{enumerate}

Firstly consider
\be
\psi = \sum_{M_{JY} \in \AAA} Y^2
+ \sum_{Q > 0}\, \sum_{M_{JY} \in \RRR_Q} Y^2 
- \sum_{Q > 0}\, \sum_{M_{JY} \in \RRR_Q} Q Y 
+ \sum_{Q > 0}\, \sum_{M_{JY} \in \SSS_Q} Y^2 
+ \sum_{Q > 0}\, \sum_{M_{JY} \in \SSS_Q} Q Y .
\ee
Take a multiplet of type~\ref{type1}.
It contributes to $\psi$ as
\bs
\ba
& & \left( 0 \right) + \left( 0 \right) - \left( 0 \right)
+ \left[ \left( - Y + J \right) - \left( - Y - J \right) + 1 \right] Y^2
\no & &
+ \left[ \left( - Y - J \right) + \cdots + \left( - Y + J \right) \right] Y
\\ &=& \left( 2 J + 1 \right) Y^2
+ \frac{\left( - Y + J \right) \left( - Y + J + 1 \right)
  - \left( - Y - J - 1 \right) \left( - Y - J \right)}{2}\ Y
\\ &=& 0.
\ea
\es
Take a multiplet of type~\ref{type2}.
It gives the following contribution to $\psi$:
\bs
\ba
& &
\left( Y^2 \right)
+ \left( 0 \right)
- \left( 0 \right)
+ \left( - 2 Y \right) Y^2
+ \left[ 1 + 2 + \cdots + \left( - 2Y \right) \right] Y
\\ &=& \left( - 2 Y + 1 \right) Y^2
+ \frac{\left( - 2 Y \right) \left( - 2 Y + 1 \right)}{2}\ Y
\\ &=& 0.
\ea
\es
Take a multiplet of type~\ref{type3}.
Its contribution to $\psi$ is
\bs
\ba
& &
Y^2 + \left( J + Y \right) Y^2
- \left[ 1 + \cdots + \left( J + Y \right) \right] Y
\\ & &
+ \left( J - Y \right) Y^2
+ \left[ 1 + \cdots + \left( J - Y \right) \right] Y
\\ &=&
\left( 2 J + 1 \right) Y^2
+ \frac{\left( J - Y \right) \left( J - Y + 1 \right)
  - \left( J + Y \right) \left( J + Y + 1 \right)}{2}\ Y
\\ &=& 0.
\ea
\es
Take a multiplet of type~\ref{type4}.
It contributes to $\psi$ as
\bs
\ba
& &
\left( Y^2 \right)
+ \left( 2 Y \right) Y^2
- \left( 1 + 2 + \cdots + 2Y \right) Y
+ \left( 0 \right)
+ \left( 0 \right)
\\ &=& \left( 2 Y + 1 \right) Y^2
- \frac{\left( 2 Y \right) \left( 2 Y + 1 \right)}{2}\ Y
\\ &=& 0.
\ea
\es
Take a multiplet of type~\ref{type5}.
Its contribution to $\psi$ is
\bs
\ba
& &
\left( 0 \right)
+ \left[ \left( Y + J \right) - \left( Y - J \right) + 1 \right] Y^2
\no & &
- \left[ \left( Y - J \right) + \cdots + \left( Y + J \right) \right] Y
+ \left( 0 \right) + \left( 0 \right)
\\ &=& \left( 2 J + 1 \right) Y^2
- \frac{\left( Y + J \right) \left( Y + J + 1 \right)
- \left( Y - J - 1 \right) \left( Y - J \right)}{2}\ Y
\\ &=& 0.
\ea
\es
We conclude that no multiplet produces a nonzero contribution to $\psi$,
hence $\psi$ is zero.

We next consider
\ba
\theta &=&
2 \sum_{M_{JY} \in \AAA} \left( J^2 + J - 2 Y^2 \right)
\\ & &
+ \sum_{Q > 0}\, \sum_{M_{JY} \in \RRR_Q}
\left[ \left( J^2 + J - 3 Y^2 + Y \right)
  + \left( 6 Y - 1 \right) Q - 3 Q^2 \right]
\\ & &
+ \sum_{Q > 0}\, \sum_{M_{JY} \in \SSS_Q}
\left[ \left( J^2 + J - 3 Y^2 - Y \right)
  - \left( 6 Y + 1 \right) Q - 3 Q^2 \right].
\ea
Take a multiplet of type~\ref{type1}.
It belongs to neither $\AAA$ nor $\RRR_Q$;
it belongs to $\SSS_Q$ from $Q = - Y - J$ to $Q = - Y + J$.
Its contribution to $\theta$ reads
\bs
\ba
& &
\left( J^2 + J - 3 Y^2 - Y \right)
\left[ \left( - Y + J \right) - \left( - Y - J \right) + 1 \right]
\\ & &
- \left( 6 Y + 1 \right)
\left[ \left( - Y - J \right) + \cdots + \left( - Y + J \right) \right]
- 3
\left[ \left( - Y - J \right)^2 + \cdots + \left( - Y + J \right)^2 \right]
\\ &=&
\left( J^2 + J - 3 Y^2 - Y \right) \left( 2 J + 1 \right)
\\ & &
- \left( 6 Y + 1 \right) \frac{\left( - Y + J \right) \left( - Y + J + 1 \right)
  - \left( - Y - J - 1 \right) \left( - Y - J \right)}{2}
\\ & &
- \frac{\left( - Y + J \right) \left( - Y + J + 1 \right)
  \left( - 2 Y + 2 J + 1 \right) - \left( - Y - J - 1 \right)
  \left( - Y - J \right) \left( - 2 Y - 2 J - 1 \right)}{2}
\no & &
\\ &=&
\left( J^2 + J - 3 Y^2 - Y \right) \left( 2 J + 1 \right)
\\ & &
- \left( 6 Y + 1 \right) \frac{\left( - Y + J \right)^2 - Y + J
  - \left( - Y - J \right)^2 - Y - J}{2}
\\ & &
- \frac{2 \left( - Y + J \right)^3 + 3 \left( - Y + J \right)^2
  - Y + J - 2 \left( - Y - J \right)^3
  + 3 \left( - Y - J \right)^2 + Y + J}{2}
\\ &=& 0.
\ea
\es
Take a multiplet of type~\ref{type2},
which has $J = - Y$.
Its contribution to $\theta$ is
\bs
\ba
& & 2 \left( Y^2 - Y - 2 Y^2 \right)
+ \left( Y^2 - Y - 3 Y^2 - Y \right) \left( - 2 Y \right)
\\ & &
- \left( 6 Y + 1 \right) \left[ 1 + 2 + \cdots + \left( - 2 Y \right) \right]
- 3 \left[ 1 + 4 + \cdots + \left( - 2 Y \right)^2 \right]
\\ &=&
- 2 Y - 2 Y^2 + 4 Y^2 + 4 Y^3
+ \left( 6 Y + 1 \right) Y \left( - 2 Y + 1 \right)
+ Y \left( - 2 Y + 1 \right) \left( - 4 Y + 1 \right)
\hspace*{7mm}
\\ &=& 0.
\ea
\es
Now consider a multiplet of type~\ref{type3}.
It produces the following contribution to $\theta$:
\bs
\ba
& & 2 \left( J^2 + J - 2 Y^2 \right)
+ \left( J^2 + J - 3 Y^2 + Y \right) \left( J + Y \right) 
+ \left( 6 Y - 1 \right)
\frac{\left( J + Y \right) \left( J + Y + 1 \right)}{2}
\hspace*{12mm} \\ & &
- \frac{\left( J + Y \right) \left( J + Y + 1 \right)
  \left( 2 J + 2 Y + 1 \right)}{2}
+ \left( J^2 + J - 3 Y^2 - Y \right) \left( J - Y \right) 
\\ & &
- \left( 6 Y + 1 \right)
\frac{\left( J - Y \right) \left( J - Y + 1 \right)}{2}
- \frac{\left( J - Y \right) \left( J - Y + 1 \right)
  \left( 2 J - 2 Y + 1 \right)}{2}
\\ &=& 0.
\ea
\es
Take a multiplet of type~\ref{type4},
\textit{i.e.}\ with $J = Y$.
It contributes to $\theta$ as
\bs
\ba
& & 2 \left( Y^2 + Y - 2 Y^2 \right)
+ \left( Y^2 + Y - 3 Y^2 + Y \right) \left( 2 Y \right)
\\ & &
+ \left( 6 Y - 1 \right) \left[ 1 + 2 + \cdots + \left( 2 Y \right) \right]
- 3 \left[ 1 + 4 + \cdots + \left( 2 Y \right)^2 \right]
\\ &=&
2 Y - 2 Y^2 + 4 Y^2 - 4 Y^3
+ \left( 6 Y - 1 \right) Y \left( 2 Y + 1 \right)
- Y \left( 2 Y + 1 \right) \left( 4 Y + 1 \right)
\\ &=& 0.
\ea
\es
Take a multiplet of type~\ref{type5},
that has only $\RRR_Q$ from $Q = Y - J$ to $Q = Y + J$.
It contributes to $\theta$ as
\bs
\ba
& &
\left( J^2 + J - 3 Y^2 + Y \right)
\left[ \left( Y + J \right) - \left( Y - J \right) + 1 \right]
\\ & &
+ \left( 6 Y - 1 \right)
\left[ \left( Y - J \right) + \cdots + \left( Y + J \right) \right]
- 3
\left[ \left( Y - J \right)^2 + \cdots + \left( Y + J \right)^2 \right]
\\ &=&
\left( J^2 + J - 3 Y^2 + Y \right) \left( 2 J + 1 \right)
\\ & &
+ \left( 6 Y - 1 \right) \frac{\left( Y + J \right) \left( Y + J + 1 \right)
  - \left( Y - J - 1 \right) \left( Y - J \right)}{2}
\\ & &
- \frac{\left( Y + J \right) \left( Y + J + 1 \right)
  \left( 2 Y + 2 J + 1 \right) - \left( Y - J - 1 \right)
  \left( Y - J \right) \left( 2 Y - 2 J - 1 \right)}{2}
\hspace*{5mm} \\ &=&
\left( J^2 + J - 3 Y^2 + Y \right) \left( 2 J + 1 \right)
\\ & &
+ \left( 6 Y - 1 \right) \frac{\left( Y + J \right)^2 + Y + J
  - \left( Y - J \right)^2  + Y - J}{2}
\\ & &
- \frac{2 \left( Y + J \right)^3 + 3 \left( Y + J \right)^2
  + Y + J - 2 \left( Y - J \right)^3
  + 3 \left( Y - J \right)^2 - Y + J}{2}
\\ &=& 0.
\ea
\es
Thus,
all the multiplets produce a null contribution to $\theta$,
which means that $\theta = 0$.

\section{The Goldstone bosons}
\label{sec:appendixGoldstone}

The terms that describe the mixing of the gauge bosons with the scalars
allow one to obtain the form of the Goldstone bosons.
For the mixing of the $W$ boson with the charged scalars we get
\bs
\label{gjr9tr}
\ba
\mathcal{L}_{W^\pm G^\mp} &=& i g\, W_\mu^- \left[
  \sum_{M_{JY} \in \AAA \cap \RRR_1} v_{JY}^\ast\,
  \sqrt{\frac{\left( J + Y \right) \left( J - Y + 1 \right)}{2}}\
  \partial^\mu M_{JY}^1
  \right. \\  & & \left.
  - \sum_{M_{JY} \in \AAA \cap \SSS_1} v_{JY}\,
  \sqrt{\frac{\left( J + Y + 1 \right) \left( J - Y \right)}{2}}\
  \partial^\mu \left( M_{JY}^{-1} \right)^\ast
  \right] + \mathrm{H.c.}
\ea
\es
We identify the quantity in Eq.~\eqref{gjr9tr} as being equal to
$i m_W W_\mu^- \partial^\mu G^+ + \mathrm{H.c.}$~\cite{Branco1999}.
This implicitly defines the phase of
\bs
\label{eq:chargedG}
\ba
G^+ &=& \frac{g}{m_W} \left[
  \sum_{M_{JY} \in \RRR_1} v_{JY}^\ast\,
  \sqrt{\frac{\left( J + Y \right) \left( J - Y + 1 \right)}{2}}\ M_{JY}^1
  \right. \\ & & \left.
  - \sum_{M_{JY} \in \SSS_1} v_{JY}\,
  \sqrt{\frac{\left( J + Y + 1 \right) \left( J - Y \right)}{2}}\
  \left( M_{JY}^{-1} \right)^\ast
  \right],
\ea
\es
where we have used
\be
\label{ugd000}
M_{JY} \in \RRR_1 \setminus \AAA \Rightarrow J - Y + 1 = 0,
\quad \quad
M_{JY} \in \SSS_1 \setminus \AAA \Rightarrow J + Y + 1 = 0.
\ee
It follows from Eq.~\eqref{eq:chargedG} that
the row matrices defined in Eqs.~\eqref{8594543} and~\eqref{85945432}
have their first matrix elements given by
\be
\label{r11s11}
\left( R^1_{JY} \right)_{11} =
\frac{g v_{JY}}{m_W}\,
\sqrt{\frac{J^2 - Y^2 + J + Y}{2}},
\qquad
\left( S^1_{JY} \right)_{11} =
- \frac{g v_{JY}^\ast}{m_W}\,
\sqrt{\frac{J^2 - Y^2 + J - Y}{2}}.
\ee
For the mixing of the $Z$ boson with the neutral scalars we get
\begin{equation}
\mathcal{L}_{Z G^0}
=  i\, \frac{g}{c_W}\, Z_\mu \sum_{M_{JY} \in \AAA}
Y\, v_{JY}\,
\partial^\mu \left( M_{JY}^0 \right)^\ast + \mathrm{H.c.}
\end{equation}
We identify this as being equal to
$m_Z Z_\mu \partial^\mu G^0 + \mathrm{H.c.}$~\cite{Branco1999}.
This implicitly defines the sign of
\be
\label{eq:neutralG}
G^0 = \frac{i g}{c_W m_Z} \sum_{M_{JY} \in \AAA} Y
\left[ v_{JY} \left( M^0_{JY} \right)^\ast - v_{JY}^\ast\, M^0_{JY} \right].
\ee
Therefore,
the row matrices defined in Eq.~\eqref{ure954} have their first matrix elements
given by
\begin{equation} 
\left( A_{JY} \right)_{11} = \frac{- \sqrt{2} g}{c_W m_Z}\
Y\, \mathrm{Im}\, v_{JY},
\qquad
\left( B_{JY} \right)_{11} = \frac{\sqrt{2} g}{c_W m_Z}\
Y\, \mathrm{Re}\, v_{JY}.
\end{equation}

We shall next demonstrate the Feynman rules in Eqs.~\eqref{GBvertices}.
Using the result of Eq.~\eqref{bugf954},
the $Z G^+ G^{-}$ interaction terms are
\bs
\ba
\mathcal{L}_{Z G^+ G^{-}} &=& 
Z^\mu \left( G^+\, \partial_\mu G^{-} - G^{-}\, \partial_\mu G^+ \right)
\left\{ - i g c_W
+ i\, \frac{g}{c_W} \left[
\sum_{M_{JY} \in \RRR_1} Y \left|\left(R_{JY}^1\right)_{1 1} \right|^2 
\right. \right. \no & & \left. \left.
- \sum_{M_{JY} \in \SSS_1} Y \left|\left( S_{JY}^1 \right)_{1 1} \right|^2
\right] \right\}
\\ &=&
Z^\mu \left( G^+\, \partial_\mu G^{-} - G^{-}\, \partial_\mu G^+ \right)
\left\{ - i g c_W
\right. \no & &
+ i\, \frac{g^3}{2 m_W^2 c_W} \left[
  \sum_{M_{JY} \in \RRR_1} \left|v_{JY}\right|^2 Y \left( J^2 - Y^2 + J + Y \right)
  \right. \no & & \left. \left.
  - \sum_{M_{JY} \in \SSS_1} \left|v_{JY}\right|^2
  Y  \left( J^2 - Y^2 + J - Y \right)
  \right] \right\}
\\ &=&
Z^\mu \left( G^+\, \partial_\mu G^{-} - G^{-}\, \partial_\mu G^+ \right)
\left\{ - i g c_W
\right. \no & &
+ i\, \frac{g^3}{2 m_W^2 c_W} \left[
  \sum_{M_{JY} \in \AAA} \left|v_{JY}\right|^2 Y \left( J^2 - Y^2 + J + Y \right)
  \right. \no & & \left. \left.
- \sum_{M_{JY} \in \AAA} \left|v_{JY}\right|^2  Y  \left( J^2 - Y^2 + J - Y \right)
\right] \right\}
\\ &=& Z^\mu \left( G^+\, \partial_\mu G^{-}
- G^{-}\, \partial_\mu G^+ \right) \left( - i g c_W
+ i\, \frac{g^3}{m_W^2 c_W}\,
\sum_{M_{JY} \in \AAA} \left| v_{JY} \right|^2 Y^2 \right)
\\ &=&
i g c_W \left( \frac{m_Z^2}{2 m_W^2} - 1 \right)
Z^\mu \left( G^+\, \partial_\mu G^{-} - G^{-}\, \partial_\mu G^+ \right),
\label{jgur555}
\ea
\es
where we have used firstly Eq.~\eqref{r11s11},
secondly Eqs.~\eqref{setminus1} and~\eqref{ugd000},
and thirdly Eq.~\eqref{mz}.
Equation~\eqref{jgur555} proves that
the Feynman rule in Eq.~\eqref{eq:ZGGvertSM} is correct
for a model with an arbitrary scalar content.

Using the result of Eq.~\eqref{pri54},
the $W^\pm G^{\mp} G^{0}$ interaction terms are 
\bs
\ba
\mathcal{L}_{W^\pm G^\mp G^0} &=&
i\, \frac{g}{2}\, \sum_{M_{JY} \in \AAA}\,
\sqrt{\left( J + Y + 1 \right) \left( J - Y \right)}
\\*[1mm] & & \times
\left[ W^{\mu +} \left( A_{JY} - i B_{JY} \right)_{11}
  \left( S_{JY}^1 \right)^\ast_{11}
  \left( G^0\, \partial_\mu G^- - G^-\, \partial_\mu G^0 \right)
  \right. \\*[0.5mm] & & \left.
  + W^{\mu -} \left( A_{JY} + i B_{JY} \right)_{11} \left( S_{JY}^1 \right)_{11}
  \left( G^+\, \partial_\mu G^0 - G^0\, \partial_\mu G^+ \right)
  \right]
\\ & & + i\, \frac{g}{2}\, \sum_{M_{JY} \in \AAA}\,
\sqrt{\left( J + Y \right) \left( J - Y + 1 \right)}
\\*[1mm] & & \times
\left[ W^{\mu +} \left( A_{JY} + i B_{JY} \right)_{11}
  \left( R_{JY}^1 \right)^\ast_{11}
  \left( G^-\, \partial_\mu G^0  - G^0\, \partial_\mu G^- \right)
  \right. \\*[0.5mm] & & \left.
  + W^{\mu -} \left( A_{JY} - i B_{JY} \right)_{11} \left( R_{JY}^1 \right)_{11}
  \left (G^0\, \partial_\mu G^+  - G^+\, \partial_\mu G^0 \right)
  \right] \\
  &=&
- \frac{g^3}{2 m_Z m_W c_W}\, \sum_{M_{JY} \in \AAA}
\left| v_{JY} \right|^2 Y \left( J + Y + 1 \right) \left( J - Y \right)
\\*[1mm] & & \times
\left[ W^{\mu +}
  \left( G^0\, \partial_\mu G^- - G^-\, \partial_\mu G^0 \right)
  - W^{\mu -} 
  \left( G^+\, \partial_\mu G^0 - G^0\, \partial_\mu G^+ \right)
  \right]
\\ & & + \frac{g^3}{2 m_Z m_W c_W}\, \sum_{M_{JY} \in \AAA}\,
\left| v_{JY} \right|^2 Y \left( J + Y \right) \left( J - Y + 1 \right)
\\*[1mm] & & \times
\left[ W^{\mu +}
  \left( G^0\, \partial_\mu G^- - G^-\, \partial_\mu G^0 \right)
  - W^{\mu -} 
  \left( G^+\, \partial_\mu G^0 - G^0\, \partial_\mu G^+ \right)
  \right]
\\ &=&
\frac{g^3}{m_Z m_W c_W}\, \sum_{M_{JY} \in \AAA} \left| v_{JY} \right|^2 Y^2
\\ & & \times
\left[ W^{\mu +} \left( G^0\, \partial_\mu G^- - G^-\, \partial_\mu G^0 \right)
+ W^{\mu -} \left( G^0\, \partial_\mu G^+ - G^+\, \partial_\mu G^0 \right) \right]
\\*[1mm] &=&
\frac{g m_Z c_W}{2 m_W}
\left[ W^{\mu +}
  \left( G^0\, \partial_\mu G^- - G^-\, \partial_\mu G^0 \right)
  + W^{\mu -} \left( G^0\, \partial_\mu G^+ - G^+\, \partial_\mu G^0 \right)
  \right]. \hspace*{7mm}
\ea
\es
This proves that the Feynman rule in Eq.~\eqref{eq:WG0GvertSM}
is correct for a model with an arbitrary scalar content.

From Eq.~\eqref{viw00e},
\bs
\ba
D_\mu M_{JY}^{-1} &=& \partial_\mu M_{JY}^{-1} - i e A_\mu M_{JY}^{-1}
+ i\, \frac{g}{c_W}\, Z_\mu M_{JY}^{-1} \left( Y + c_W^2 \right)
\\ & &
- i g\, W_\mu^+ M_{JY}^{-2}\,
\sqrt{\frac{\left( J + Y + 2 \right) \left( J - Y - 1 \right)}{2}}
\\ & &
- i g\, W_\mu^- M_{JY}^{0}\,
\sqrt{\frac{\left( J + Y + 1 \right) \left( J - Y \right)}{2}},
\\
D_\mu M_{JY}^0 &=& \partial_\mu M_{JY}^0 + i\, \frac{g}{c_W}\, Z_\mu M_{JY}^0 Y
\\ & &
- i g\, W_\mu^+ M_{JY}^{-1}\,
\sqrt{\frac{\left( J + Y + 1 \right) \left( J - Y \right)}{2}}
\\ & &
- i g\, W_\mu^- M_{JY}^1\,
\sqrt{\frac{\left( J + Y \right) \left( J - Y + 1 \right)}{2}},
\\
D_\mu M_{JY}^1 &=& \partial_\mu M_{JY}^1 + i e A_\mu M_{JY}^1
+ i\, \frac{g}{c_W}\, Z_\mu M_{JY}^1 \left( Y - c_W^2 \right)
\\ & &
- i g\, W_\mu^+ M_{JY}^0\,
\sqrt{\frac{\left( J + Y \right) \left( J - Y + 1 \right)}{2}}
\\ & &
- i g\, W_\mu^- M_{JY}^2\,
\sqrt{\frac{\left( J + Y - 1 \right) \left( J - Y + 2 \right)}{2}}.
\ea
\es
Therefore,
the interaction terms of the $Z^0$ with $W^\pm$
and a charged scalar are given by
\bs
\label{eq:ZWS}
\ba
\mathcal{L}_{Z W^\pm S^{\mp}} &=&
- \frac{g^2}{c_W}\, Z^\mu W^{+}_\mu \left[
  \sum_{M_{JY} \in \AAA \cap \SSS_1} \left( Y + c_W^2 \right)
  \sqrt{\frac{\left( J + Y + 1 \right) \left( J - Y \right)}{2}}\
  v_{JY}^\ast M_{JY}^{-1}
  \right. \\ & & \left.
  + \sum_{M_{JY} \in \AAA \cap \SSS_1}
  Y\, \sqrt{\frac{\left( J + Y + 1 \right) \left( J - Y \right)}{2}}\
  v_{JY}^\ast M_{JY}^{-1} \right]
\\ & &
- \frac{g^2}{c_W}\, Z_\mu  W^-_\mu \left[
  \sum_{M_{JY} \in \AAA \cap \RRR_1}
  Y\, \sqrt{\frac{\left( J + Y  \right) \left( J - Y + 1 \right)}{2}}\
  v_{JY}^\ast M_{JY}^1
  \right. \\ & & \left.
  + \sum_{M_{JY} \in \AAA \cap \RRR_1} \left( Y - c_W^2 \right)
  \sqrt{\frac{\left( J + Y  \right) \left( J - Y + 1 \right)}{2}}\
  v_{JY}^\ast M_{JY}^1 \right]
+ \mathrm{H.c.}
\\ &=&
- \frac{g^2}{c_W}\, Z^\mu W^+_\mu
\sum_{M_{JY} \in \AAA \cap \SSS_1} \left( 2Y + c_W^2 \right)
\sqrt{\frac{\left( J + Y + 1 \right) \left( J - Y \right)}{2}}\
v_{JY}^\ast M_{JY}^{-1} 
\\ & &
- \frac{g^2}{c_W}\, Z_\mu W^-_\mu
\sum_{M_{JY} \in \AAA \cap \RRR_1} \left( 2Y - c_W^2 \right)
\sqrt{\frac{\left( J + Y  \right) \left( J - Y + 1 \right)}{2}}\
v_{JY}^\ast M_{JY}^1
\\ & &
+ \mathrm{H.c.}
\\ &=&
\frac{g m_W}{c_W}\, Z^\mu  W^+_\mu \! \sum_{M_{JY} \in \AAA \cap \SSS_1}
\left( 2Y + c_W^2 \right) \left( S_{JY} \right)_{11}\,
\sum_{a=1}^{n_1} \left( S_{JY}^{1} \right)^\ast_{1a} S_a^-
\\ & &
- \frac{g m_W}{c_W}\, Z_\mu  W^-_\mu \! \sum_{M_{JY} \in \AAA \cap \RRR_1}
\left( 2Y - c_W^2 \right) \left( R_{JY} \right)_{11}^\ast\,
\sum_{a=1}^{n_1} \left( R_{JY}^1 \right)_{1a} S_a^+
\\ & &
+ \mathrm{H.c.}
\\ &=&
\frac{g m_W}{c_W}\, Z_\mu W^{\mu -}
\sum_{M_{JY} \in \mathcal{A} \cap \mathcal{S}_1}
\left( 2Y + c_W^2 \right) \left( S_{JY}^1 \right)^\ast_{11}\,
\sum_{a=1}^{n_1} \left( S_{JY}^{1} \right)_{1 a} S_a^+
\\ & &
- \frac{g m_W}{c_W}\, Z_\mu W^{\mu -} \sum_{M_{JY} \in \mathcal{A} \cap \mathcal{R}_1}
\left( 2Y - c_W^2 \right) \left( R_{JY}^1 \right)^\ast_{11}\,
\sum_{a=1}^{n_1} \left( R_{JY}^1 \right)_{1 a} S_a^+
\\ & &
+ \mathrm{H.c.}
\ea
\es
Therefore,
the $Z W^\pm G^{\mp}$ interaction terms are
\bs
\ba
\mathcal{L}_{Z W^\pm G^{\mp}} \! &=& \!
\frac{g m_W}{c_W}\, Z_\mu \left[
  \sum_{M_{JY} \in \mathcal{A} \cap \mathcal{S}_{1}}
  \left| \left( S_{JY}^1 \right)_{11} \right|^2 \left( 2Y + c_W^2 \right)
  \right. \\ & & \left.
  - \sum_{M_{JY} \in \mathcal{A} \cap \mathcal{R}_{1}}
  \left| \left( R_{JY}^1 \right)_{11} \right|^2 \left( 2Y - c_W^2 \right)
  \right]
\left( W^{\mu -} G^+ + W^{\mu +} G^- \right)
\\ &=&
\frac{g^3}{2 m_W c_W}\, Z_\mu W^{\mu -} G^+ \left[
  \sum_{M_{JY} \in \mathcal{A} \cap \mathcal{S}_{1}}
  \left| v_{JY} \right|^2 \left( J^2 - Y^2 + J - Y \right)
  \left( 2Y + c_W^2 \right)
  \right. \\ & & \left.
  - \sum_{M_{JY} \in \mathcal{A} \cap \mathcal{R}_{1}}
  \left| v_{JY} \right|^2 \left( J^2 - Y^2 + J + Y \right)
  \left( 2Y - c_W^2 \right)
  \right]
+ \mathrm{H.c.}
\\ &=&
\frac{g^3}{2 m_W c_W}\, Z_\mu W^{\mu -} G^+ \left[
  \sum_{M_{JY} \in \mathcal{A}} \left| v_{JY} \right|^2
  \left( J^2 - Y^2 + J - Y \right) \left( 2Y + c_W^2 \right)
  \right. \\ & & \left.
  - \sum_{M_{JY} \in \mathcal{A}} \left| v_{JY} \right|^2
  \left( J^2 - Y^2 + J + Y \right) \left( 2Y - c_W^2 \right)
  \right]
+ \mathrm{H.c.}
\\ &=&
\frac{g^3}{m_W c_W}\, Z_\mu W^{\mu -} G^+
\sum_{M_{JY} \in \mathcal{A}} \left| v_{JY} \right|^2
\left[ - 2 Y^2 + c_W^2 \left( J^2 - Y^2 + J \right) \right]
+ \mathrm{H.c.} \hspace*{7mm}
\\ &=&
\frac{g}{m_W c_W}\, Z_\mu W^{\mu -} G^+
\left( - c_W^2 m_Z^2 + c_W^2 m_W^2 \right)
+ \mathrm{H.c.}
\ea
\es
This proves the Feynman rule in Eq.~\eqref{eq:ZWGvertSM}
for a model with an arbitrary scalar content.

\section{The Feynman diagrams that produce
  the parameter \texorpdfstring{$S$}{S} in a model with triplets}
\label{sec:diagrams}

In order to compute the oblique parameter $S$
we need to compute the vacuum-polarization tensors $\Pi_{VV^\prime}^{\mu \nu}$,
where $V$ and $V^\prime$ may be either $A$ and $A$
($A$ is the photon),
$A$ and $Z$,
or $Z$ and $Z$.

The diagrams that contribute to $\left. \frac{\partial \, \delta A_{Z Z}(q^2)}{\partial q^2} \right|_{q^2=0}$
at the one-loop level are\footnote{Here and below
we do not include the diagrams that correspond to the same amplitudes
in the model with triplets as in the Standard Model,
since those diagrams do not contribute to $S$.}
\begin{subequations} \label{eq:ZZ}
\allowdisplaybreaks
\begin{align}
\parbox{40mm}{
\begin{fmfgraph*}(100,100)
    \fmfleft{i1}
    \fmfright{o1}
    \fmf{boson,label=$Z$,label.side=left}{i1,w1}
    \fmf{boson,label=$Z$,label.side=left}{w2,o1}
    \fmf{scalar,right,tension=.3,label=$S^{--}_{c^\prime}$}{w1,w2}
    \fmf{scalar,left,tension=.3,label=$S^{++}_c$}{w1,w2}
    \fmflabel{$\mu$}{i1}
    \fmflabel{$\nu$}{o1}
    \marrow{a}{down}{bot}{$q$}{i1,w1}{(1/4, 2/3)}
    \marrow{b}{down}{bot}{$q$}{w2,o1}{(1/3, 3/4)}
\end{fmfgraph*}} & \qquad \qquad \qquad 
\parbox{40mm}{
\begin{fmfgraph*}(100,100)
    \fmfleft{i1}
    \fmfright{o1}
    \fmf{boson,label=$Z$,label.side=left}{i1,w1}
    \fmf{boson,label=$Z$,label.side=left}{w2,o1}
    \fmf{scalar,right,tension=.3,label=$S^-_{a^\prime}$}{w1,w2}
    \fmf{scalar,left,tension=.3,label=$S^+_a$}{w1,w2}
    \fmflabel{$\mu$}{i1}
    \fmflabel{$\nu$}{o1}
    \marrow{a}{down}{bot}{$q$}{i1,w1}{(1/4, 2/3)}
    \marrow{b}{down}{bot}{$q$}{w2,o1}{(1/3, 3/4)}
\end{fmfgraph*}} \\
\parbox{40mm}{
\begin{fmfgraph*}(100,100)
    \fmfleft{i1}
    \fmfright{o1}
    \fmf{boson,label=$Z$,label.side=left}{i1,w1}
    \fmf{boson,label=$Z$,label.side=left}{w2,o1}
    \fmf{dashes,right,tension=.3,label=$S^0_{b^\prime}$}{w1,w2}
    \fmf{dashes,left,tension=.3,label=$S^0_b$}{w1,w2}
    \fmflabel{$\mu$}{i1}
    \fmflabel{$\nu$}{o1}
    \marrow{a}{down}{bot}{$q$}{i1,w1}{(1/4, 2/3)}
    \marrow{b}{down}{bot}{$q$}{w2,o1}{(1/3, 3/4)}
\end{fmfgraph*}} & \qquad \qquad \qquad
 \parbox{40mm}{
\begin{fmfgraph*}(100,100)
    \fmfleft{i1}
    \fmfright{o1}
    \fmf{boson,label=$Z$,label.side=left}{i1,w1}
    \fmf{boson,label=$Z$,label.side=left}{w2,o1}
    \fmf{wiggly_arrow,right,tension=.3,label=$W^-$}{w1,w2}
    \fmf{scalar,left,tension=.3,label=$S^+_a$}{w1,w2}
    \fmflabel{$\mu$}{i1}
    \fmflabel{$\nu$}{o1}
    \marrow{a}{down}{bot}{$q$}{i1,w1}{(1/4, 2/3)}
    \marrow{b}{down}{bot}{$q$}{w2,o1}{(1/3, 3/4)}
\end{fmfgraph*}}  \\
\parbox{40mm}{
\begin{fmfgraph*}(100,100)
    \fmfleft{i1}
    \fmfright{o1}
    \fmf{boson,label=$Z$,label.side=left}{i1,w1}
    \fmf{boson,label=$Z$,label.side=left}{w2,o1}
    \fmf{wiggly_arrow,right,tension=.3,label=$W^+$}{w1,w2}
    \fmf{scalar,left,tension=.3,label=$S^-_a$}{w1,w2}
    \fmflabel{$\mu$}{i1}
    \fmflabel{$\nu$}{o1}
    \marrow{a}{down}{bot}{$q$}{i1,w1}{(1/4, 2/3)}
    \marrow{b}{down}{bot}{$q$}{w2,o1}{(1/3, 3/4)}
\end{fmfgraph*}} & \qquad \qquad \qquad 
\parbox{40mm}{
\begin{fmfgraph*}(100,100)
    \fmfleft{i1}
    \fmfright{o1}
    \fmf{boson,label=$Z$,label.side=left}{i1,w1}
    \fmf{boson,label=$Z$,label.side=left}{w2,o1}
    \fmf{wiggly,right,tension=.3,label=$Z$}{w1,w2}
    \fmf{dashes,left,tension=.3,label=$S^0_b$}{w1,w2}
    \fmflabel{$\mu$}{i1}
    \fmflabel{$\nu$}{o1}
    \marrow{a}{down}{bot}{$q$}{i1,w1}{(1/4, 2/3)}
    \marrow{b}{down}{bot}{$q$}{w2,o1}{(1/3, 3/4)}
\end{fmfgraph*}}
\end{align}
\end{subequations}
The diagrams that contribute to $\left. \frac{\partial \, \delta A_{A A}(q^2)}{\partial q^2} \right|_{q^2=0}$ at one-loop level are 
\begin{subequations} \label{eq:AA}
\allowdisplaybreaks
\begin{align}
\parbox{40mm}{
\begin{fmfgraph*}(100,100)
    \fmfleft{i1}
    \fmfright{o1}
    \fmf{boson,label=$A$,label.side=left}{i1,w1}
    \fmf{boson,label=$A$,label.side=left}{w2,o1}
    \fmf{scalar,right,tension=.3,label=$S^{--}_{c}$}{w1,w2}
    \fmf{scalar,left,tension=.3,label=$S^{++}_c$}{w1,w2}
    \fmflabel{$\mu$}{i1}
    \fmflabel{$\nu$}{o1}
    \marrow{a}{down}{bot}{$q$}{i1,w1}{(1/4, 2/3)}
    \marrow{b}{down}{bot}{$q$}{w2,o1}{(1/3, 3/4)}
\end{fmfgraph*}} & \qquad \qquad \qquad 
\parbox{40mm}{
\begin{fmfgraph*}(100,100)
    \fmfleft{i1}
    \fmfright{o1}
    \fmf{boson,label=$A$,label.side=left}{i1,w1}
    \fmf{boson,label=$A$,label.side=left}{w2,o1}
    \fmf{scalar,right,tension=.3,label=$S^-_{a}$}{w1,w2}
    \fmf{scalar,left,tension=.3,label=$S^+_a$}{w1,w2}
    \fmflabel{$\mu$}{i1}
    \fmflabel{$\nu$}{o1}
    \marrow{a}{down}{bot}{$q$}{i1,w1}{(1/4, 2/3)}
    \marrow{b}{down}{bot}{$q$}{w2,o1}{(1/3, 3/4)}
\end{fmfgraph*}} 
\end{align}
\end{subequations}
The diagrams that contribute to $\left. \frac{\partial \, \delta A_{A Z}(q^2)}{\partial q^2} \right|_{q^2=0}$ at one-loop level are
\begin{subequations} \label{eq:AZ}
\allowdisplaybreaks
\begin{align}
\parbox{40mm}{
\begin{fmfgraph*}(100,100)
    \fmfleft{i1}
    \fmfright{o1}
    \fmf{boson,label=$A$,label.side=left}{i1,w1}
    \fmf{boson,label=$Z$,label.side=left}{w2,o1}
    \fmf{scalar,right,tension=.3,label=$S^{--}_{c}$}{w1,w2}
    \fmf{scalar,left,tension=.3,label=$S^{++}_c$}{w1,w2}
    \fmflabel{$\mu$}{i1}
    \fmflabel{$\nu$}{o1}
    \marrow{a}{down}{bot}{$q$}{i1,w1}{(1/4, 2/3)}
    \marrow{b}{down}{bot}{$q$}{w2,o1}{(1/3, 3/4)}
\end{fmfgraph*}} & \qquad \qquad \qquad 
\parbox{40mm}{
\begin{fmfgraph*}(100,100)
    \fmfleft{i1}
    \fmfright{o1}
    \fmf{boson,label=$A$,label.side=left}{i1,w1}
    \fmf{boson,label=$Z$,label.side=left}{w2,o1}
    \fmf{scalar,right,tension=.3,label=$S^-_{a}$}{w1,w2}
    \fmf{scalar,left,tension=.3,label=$S^+_a$}{w1,w2}
    \fmflabel{$\mu$}{i1}
    \fmflabel{$\nu$}{o1}
    \marrow{a}{down}{bot}{$q$}{i1,w1}{(1/4, 2/3)}
    \marrow{b}{down}{bot}{$q$}{w2,o1}{(1/3, 3/4)}
\end{fmfgraph*}}
\end{align}
\end{subequations}

\end{fmffile}

\end{document}